\documentclass[12pt]{article}
\usepackage{authblk}
\usepackage{lineno}
\modulolinenumbers[5]
\usepackage{amsbsy,amsmath,amsthm,latexsym,amssymb}
\usepackage{graphicx,stmaryrd,sectsty}
\usepackage{setspace}
\usepackage[font=small,labelfont=bf]{caption}
\usepackage{verbatim}
\usepackage{graphicx}
\usepackage{caption}
\usepackage{subcaption}
\usepackage{float,color}
\newcommand{\bc}{\begin{center}}
	\newcommand{\ec}{\end{center}}
\newcommand{\bfr}{\begin{flushright}}
	\newcommand{\efr}{\end{flushright}}

\newcommand{\no}{\noindent}
\newcommand{\be}{\begin{enumerate}}
	\newcommand{\ee}{\end{enumerate}}
\newcommand{\bi}{\begin{itemize}}
	\newcommand{\ei}{\end{itemize}}
\newcommand{\bd}{\begin{description}}
	\newcommand{\ed}{\end{description}}
\newcommand{\beq}{\begin{equation}}
	\newcommand{\eeq}{\end{equation}}
\newcommand{\bea}{\begin{eqnarray}}
	\newcommand{\eea}{\end{eqnarray}}

\newcommand{\bfi}{\begin{figure}}
	\newcommand{\efi}{\end{figure}}
\newcommand{\bay}{\begin{array}{l}}
	\newcommand{\eay}{\end{array}}

\newcommand{\cref}[1]{(\ref{#1})}   

\sectionfont{\normalsize}
\subsectionfont{\normalsize}
\subsubsectionfont{\normalsize}
\usepackage{siunitx,tipa,booktabs}
\begin{document}
	
	
	\begin{titlepage}
		\clearpage\thispagestyle{empty}
		\noindent
		\hrulefill
		\begin{figure}[h!]
			\centering
			\includegraphics[width=2 in]{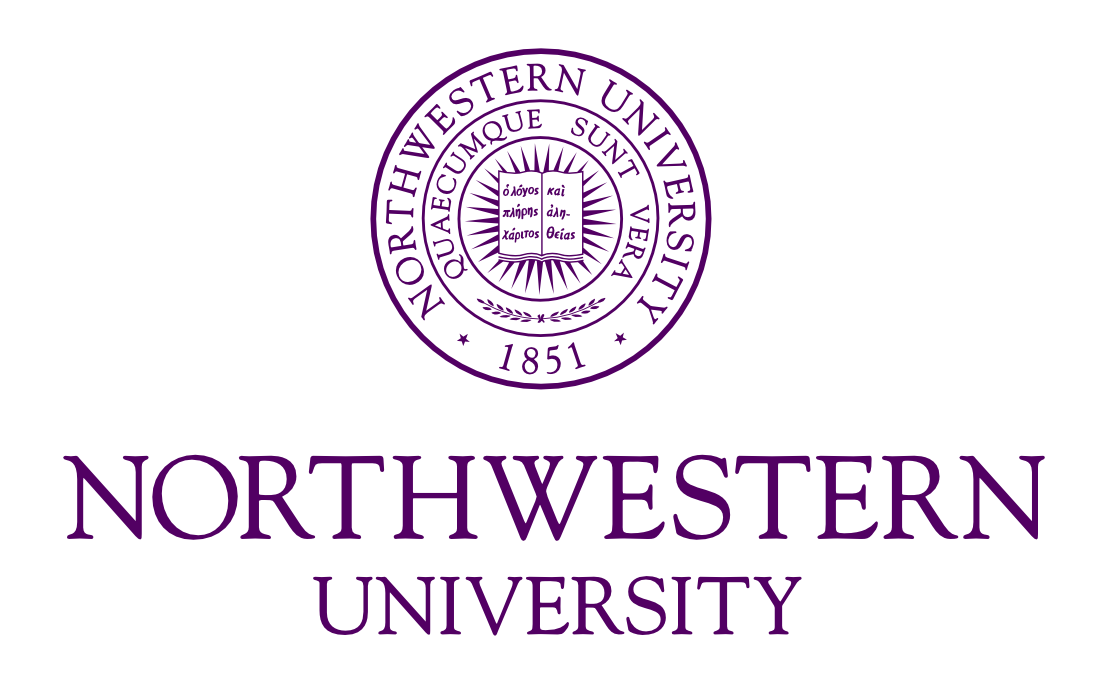}
		\end{figure}
		\begin{center}
			{
				{
					{\bf Center for Sustainable Engineering of Geological and Infrastructure Materials} \\ [0.1in]
					Department of Civil and Environmental Engineering \\ [0.1in]
					McCormick School of Engineering and Applied Science \\ [0.1in]
					Evanston, Illinois 60208, USA
				}
			}
		\end{center} 
		\hrulefill \\ \vskip 2mm
		\vskip 0.5in
		\begin{center}
			{\large {\bf A MULTISCALE FRAMEWORK FOR THE SIMULATION OF THE ANISOTROPIC MECHANICAL BEHAVIOR OF SHALE
			}}\\[0.5in]
			{\large {\sc Weixin Li, Roozbeh Rezakhani, Congrui Jin, Xinwei Zhou, Gianluca Cusatis}}\\[0.75in]
			{\sf \bf SEGIM INTERNAL REPORT No. 16-8/478M}\\[0.75in]
		\end{center}
		\vskip 5mm
	\noindent {\footnotesize {{\em Submitted to International Journal for Numerical and Analytical Methods in Geomechanics \hfill August 2016} }}
	\end{titlepage}

	\newpage
	\clearpage \pagestyle{plain} \setcounter{page}{1}
	\title{A multiscale framework for the simulation of the anisotropic mechanical behavior of shale}
	\author[1]{\small Weixin Li}
	\author[2]{Roozbeh Rezakhani\thanks{Corresponding author}}
	\author[3]{Congrui Jin}
	\author[4]{Xinwei Zhou}
	\author[1,2]{Gianluca Cusatis}
	
	\affil[1]{\footnotesize Theoretical and Applied Mechanics, Northwestern University, Evanston, IL 60208, U.S.A.}
	\affil[2]{\footnotesize Department of Civil and Environmental Engineering, Northwestern University, Evanston, IL 60208, U.S.A.}	
	\affil[3]{\footnotesize Department of Mechanical Engineering, State University of New York at Binghamton, Binghamton, NY 13902, U.S.A.}	
	\affil[4]{\footnotesize ES3, 550 West C St., San Diego, CA 92101, U.S.A}		

	\date{}
	\maketitle
	\let\thefootnote\relax\footnote{\footnotesize \textit{Email addresses:} w.li@u.northwestern.edu (Weixin Li), roozbehrezakhani2011@u.northwestern.edu (Roozbeh Rezakhani), cjin@binghamton.edu (Congrui Jin), xinwei.zhou@es3inc.com (Xinwei Zhou), g-cusatis@northwestern.edu (Gianluca Cusatis)
	}	
	{\small \no {\bf   Abstract}: 
	Shale, like many other sedimentary rocks, is typically heterogeneous, anisotropic, and is characterized by partial alignment of anisotropic clay minerals and naturally formed bedding planes. In this study, a micromechanical framework based on the Lattice Discrete Particle Model (LDPM) is formulated to capture these features. Material anisotropy is introduced through an approximated geometric description of shale internal structure, which includes representation of material property variation with orientation and explicit modeling of parallel lamination. The model  is calibrated by carrying out numerical simulations to match various experimental data, including the ones relevant to elastic properties, Brazilian tensile strength, and unconfined compressive strength. Furthermore, parametric study is performed to investigate the relationship between the mesoscale parameters and the macroscopic properties. It is shown that the dependence of the elastic stiffness, strength, and failure mode on loading orientation can be captured successfully. Finally, a homogenization approach based on the asymptotic expansion of field variables is applied to upscale the proposed micromechanical model, and the properties of the homogenized model are analyzed. 
	}
	
	
	\section{Introduction}
	
	By the rapid growth of the shale gas/oil industry, especially with the development of hydraulic fracturing techniques, deep understanding of the mechanical properties of shale-like rocks is of vital importance. Gas/oil shale, described as organic rich and fine grained \cite{eseme2007review}, exhibits significant mechanical anisotropy and heterogeneity. Developing adequate numerical models to capture these complicated characteristics of shale leads to a better clarification of the influence of material properties on induced fracture initiation, propagation, and fracture simulation. Therefore, it provides a powerful tool to predict and optimize the fracturing process. 
	
\begin{figure}
		\centering
		\includegraphics[width = 1.0\textwidth]{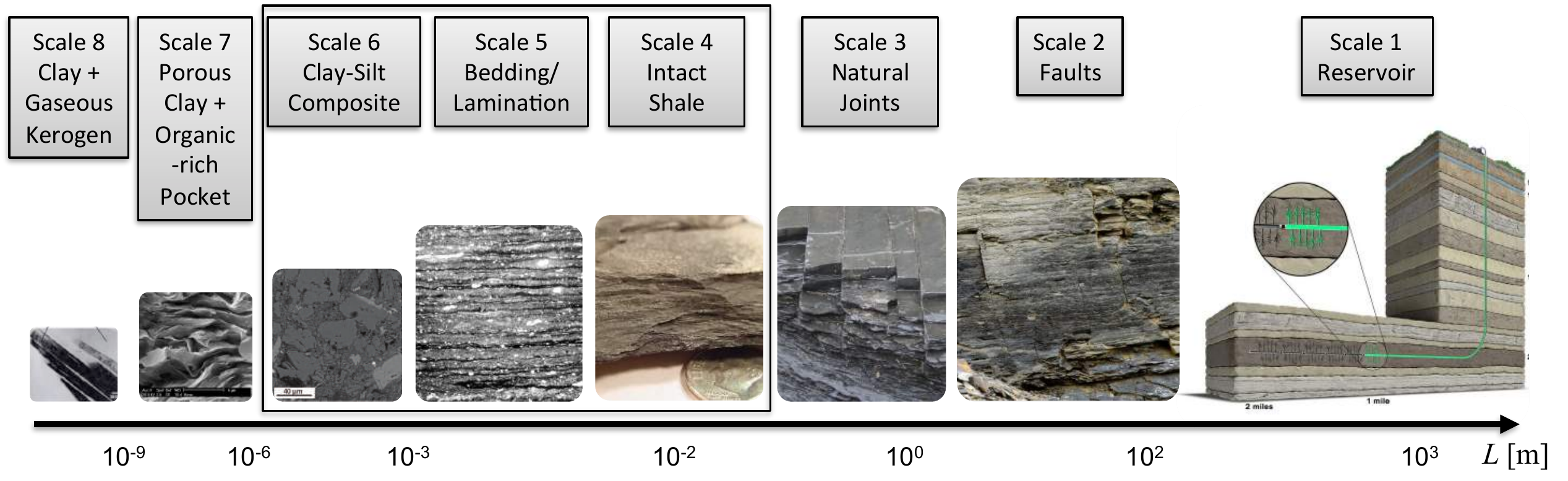}
		\caption{Multiscale structure of black shale. }
		\label{fig:MultiscaleShale}
	\end{figure}

	Shale is a highly complex and heterogeneous material that can be characterized by several levels of hierarchy as illustrated in Figure \ref{fig:MultiscaleShale}. At the microscopic level (length scale 6), it is a composite material made of porous clay, silt inclusion, and organic matter; at a lower length scale (scale 7), it exists as a porous clay/organic matter composite. The nanometer length level (scale 8) is the fundamental scale where elementary clay minerals are bounded to kerogen. At these levels, advanced measurement and experimental techniques, including scanning electron microscopy, atomic force microscopy, and nano-indentation, are widely used to characterize the morphology, topology, and mechanical properties such as elastic and poroelastic moduli \cite{bobko2008nano,abousleiman2007geomechanics,ulm2006nanogranular,bennett2015instrumented,bobko2011nanogranular}. At the macroscopic (scales 4) and mesoscopic (scale 5) levels, shale consists of a layered sedimentary rock. It is often considered as a transversely isotropic continuum, in which the anisotropy is induced by the presence of weak planes due to the sedimentation process. Hence, at this scale, shale samples exhibit mm- and \SI{}{\micro\meter}- grain-size variability \cite{slatt2011merging}, lamination/bedding planes, and stratification, which play a significant role in shale geomechanical characteristics and rock failure. At these levels, elastic, fracture, and poromechanical properties are typically accessed via standard lab measurements, including Ultrasonic Pulse Velocity (UPV), Brazilian, uniaxial compression, triaxial, and tree-point-bending tests \cite{abousleiman2007geomechanics,sone2013mechanical1,sone2013mechanical2,cho2012deformation,kim_anisotropy_2012,niandou1997laboratory}. Some new technologies such as Inclined Direct Shear Testing Device \cite{abousleiman2008laboratory} and scratch test \cite{akono2013assessment} are also developed for rock mechanics characterization at scales 4 and 5. Although faults and natural joints observed within shale formations at the field scales (scales 2-3) significantly influence the mechanical and hydraulic properties of rock masses, they are beyond the scope of this paper. In the current research, the main focus is on gas/oil shale composites at the length scales 4-6, which, in this paper, will be referred as micro-, meso-, and macro- scales, respectively. Adequate knowledge and prediction of the structure and mechanical properties at these scales are pivotal to any successful reservoir modeling endeavor. 
	
	Material heterogeneity and anisotropy are typically observed for intact shale specimens, such as Woodford shale \cite{abousleiman2007geomechanics,slatt2011merging,sondergeld2011elastic}, Barnett shale \cite{slatt2011merging,sone2013mechanical1,sone2013mechanical2}, Haynesville shale \cite{sone2013mechanical1,sone2013mechanical2,sondergeld2011elastic}, Mancos shale \cite{fjaer2014impact,simpson2014failure}, Boryeong shale  \cite{cho2012deformation,kim_anisotropy_2012,vervoort2014failure}, and Tournemire shale \cite{niandou1997laboratory}. Especially, deformation and strength anisotropy are detected in both UPV measurements \cite{sone2013mechanical1,slatt2011merging,sondergeld2011elastic}, Brazilian tensile \cite{vervoort2014failure,cho2012deformation,simpson2014failure}, uniaxial compression \cite{cho2012deformation,fjaer2014impact}, triaxial compression \cite{ambrose2014failure,niandou1997laboratory}, and triaxial creep tests \cite{sone2013mechanical1,sone2013mechanical2}. In addition, researchers reported that the heterogeneous and anisotropic nature of shale has an impact on break down pressure, fracture initiation, and fracture containment during hydraulic fracturing processes \cite{khan2012impact,suarez2006effect,higgins2008anisotropic}. Therefore, a reliable numerical model for the mechanical characterization of shale needs to take into account these properties. 
	
	As shale at the macro- and meso-scales, layered rocks are usually modeled by the numerical methods classified into continuum- and discontinuum-based approaches \cite{bobet2009numerical,jing2003review}. The continuum-based approach treats the material domain of interest as a single continuous body, and captures material failure process through commonly used techniques such as plastic softening and damage models. Spurious mesh sensitivity is the main drawback of classical continuum models, which is due to the lack of an internal length scale \cite{bobet2009numerical,lisjak2014review}. This shortcoming can be overcome by introducing micro-structural effects through second gradient damage models \cite{muhlhaus1991variational,collin2006finite}, non-local models \cite{bazant1988nonlocal}, and other high order models \cite{adhikary1997cosserat,riahi2009full} as well as crack band regularization \cite{BazOh85}. On the contrary, the discontinuum-based approach, such as discrete element method (DEM), considers the material domain as an assembly of rigid particles, and incorporates the length scale automatically. To some extend, the discrete modeling approaches are compelling when the material exhibits the lack of continuity, which makes continuum constitutive models inefficient. Hybrid approaches, such as combined finite-discrete element method (FEM/DEM) \cite{munjiza2004combined,munjiza1995combined}, are also widely used in engineering applications. Computational techniques that are used for modeling material anisotropy and layered structure are often classified into smeared and discrete approaches. The smeared approach utilizes an implicit representation of layers to produce a fictitious continuous material within which the effect of layering is introduced at the level of constitutive laws. Deformation anisotropy is commonly captured using the theory of elasticity for transversely isotropic media \cite{goodman1989introduction}, while strength anisotropy and progressive damage are captured by various anisotropic failure criteria \cite{duveau1998assessment,pietruszczak2002modelling,mahjoub2015approach,rouabhi2007triaxial,crook2002development} along with specifying the directional dependence of material properties  \cite{lisjak_continuumdiscontinuum_2014}. Discrete approaches, instead, employ an explicit representation of layers in which mechanical anisotropy is introduced by a geometric description of layered structure with varying material parameters. Strength anisotropy due to stratification or lamination, which is usually approximated by the presence of discontinuities, such as plane of weakness \cite{duveau1998assessment,jaeger1960shear,tien2001failure}, pre-existing cracks \cite{lisjak2014numerical}, parallel continuous weak layers \cite{debecker2013two}, continuous smooth joints \cite{park2013discrete, park2015bonded}, and discontinuous smooth joints \cite{duan2015discrete} models, can be captured naturally in this approach. 
	
	In the current study, the effect of the mechanical anisotropy of shale is incorporated within the framework of the Lattice Discrete Particle Model (LDPM). LDPM, developed by Cusatis and coworkers \cite{cusatis2011lattice1,cusatis2011lattice2}, is a discrete model built to mimic the micro- and meso-structure of materials. Along with its predecessor, the confinement-shear lattice model \cite{cusatis2003confinement1, cusatis2003confinement, cusatis2006confinement}, it adopts constitutive laws analogous to those of the microplane model \cite{bazant2000microplane}. LDPM has been extensive calibrated and validated against a large variety of loading conditions in both quasi-static and dynamic loading conditions, and it was demonstrated to possess superior predictive capability. Hereinafter, a formulation based on LDPM is developed to capture the mechanical anisotropy of shale. Model calibration is carried out based on the experimental data of Boryeong shale \cite{cho2012deformation,kim_anisotropy_2012}, although the model is generally applicable to other important gas shales. Finally, a homogenization approach is developed to approximate effective material characteristics, and to upscale the proposed fine scale model to enable its application to engineering practice. 
	\section{Discrete Micromechanical Modeling}
	A discrete approach based on a micromechanical framework is proposed herein, which is built upon the mesoscopic level of shale (length scale 5 as illustrated in Figure \ref{fig:MultiscaleShale}) with an explicit representation of the laminated internal structure. Main features at smaller length scales (length scale 6 and below) are tackled by the formulation of appropriate constitutive laws, while simulation of the macroscopic behavior of shale (length scale 4) is achieved by introducing a proper homogenization algorithm. The proposed micromechanical approach is formulated within the LDPM framework, which offers the following unique advantages: 
	$1)$ LDPM has been proven to be a powerful tool to accurately model the mechanical behavior of quasi-brittle materials, such as concrete and rock, under various loading conditions in both tension and unconfined, confined, and hydrostatic compression.
	$2)$ LDPM is formulated within the framework of discrete models, which provides an inherent potential to take into account the heterogeneous nature of shale. 
	$3)$ LDPM is able to explicitly reproduce the material internal structure, which provides a potential to capture the fabric anisotropy of shale. 
	Since LDPM was first proposed to simulate the failure behavior of concrete \cite{cusatis2011lattice1} and pressure-dependent inelastic processes in granular sandstone \cite{esna2015micro2} with statistically isotropic random mesostructures, it needs to be extended to capture aspects of material anisotropy based on the composition and internal texture of shale, as discussed in section \ref{sec:Geom}. The LDPM constitutive equations are introduced in section \ref{sec:Const}, and modified to accommodate material anisotropy. 
	
	\subsection{Geometrical characterization of shale internal structure}\label{sec:Geom}
	Shale at the microscopic level is often considered as a three-phase material (minerals, organic matter, and pore fill) \cite{eseme2007review,sone2013mechanical1}. The mechanical properties depend on the volume fractions of these three phases \cite{sayers2013effect,sone2013mechanical1}. The size of minerals (e.g., quartz, feldspar, rock fragments, clays, etc.) varies widely: very fine-grained particles can be smaller than \SI{2}{\micro\meter}; the silt-size particles vary from \SIrange{2}{60}{\micro\meter}; fine sand particles can be as large as \SI{60}{\micro\meter} \cite{eseme2007review, passey2010oil}. These large and small mineral grains mix with each other and tend to form boundaries where the crack appears to flow around \cite{curtis2010structural}. Organic matter can be seen in the spaces between inorganic grains or mixed with small grains \cite{curtis2010structural}. It was shown that the presence and concentration of organic matter which act as binding agents is a significant factor in the formation and stability of mud aggregates \cite{passey2010oil}. Pores, found in both organic matter, between grains, within minerals and in the form of microcracks \cite{sondergeld2010micro}, are typically in the nanometer scale, and are usually invisible at the scales under study. 
	
	The organized distribution of minerals and compliant organic materials could act as an important source of anisotropy. Sone and Zoback \cite{sone2013mechanical1} observed that fabric anisotropy forming the bedding planes depends on a combination of the following factors: preferred orientations of matrix clay, shape/orientation/distribution of organics, and alignment of elongated fossils. Sayers \cite{sayers2013effect} mentioned that anisotropy in shale results from a partial alignment of anisotropic clay particles, kerogen inclusions, bedding-parallel microcracks, low-aspect ratio pores, and layering.  The spatial distribution of shale constituents often leads to a laminated texture at the microscopic and mescoscopic levels. For example, Vernik and Nur \cite{vernik1992ultrasonic} observed the texture of organic-rich Bakken shale samples, and characterized it by bedding-parallel lamination that can be identified either in thin sections as alternating 0.2 to 3 \SI{}{mm} thick laminae or in scanning electron microscopic (SEM) backscatter images as up to 20 \SI{}{\micro\meter} thick dark laminae enriched in organic matter. In addition, Slatt et al.  \cite{slatt2011merging} suggested that laminae/bedding planes, which are the product of transport/depositional events, are planes of weakness that result in anisotropy of deformation and strength, and can affect drilling and hydraulic fracturing.  Eseme et al. \cite{eseme2007review} reported that the micro-lamination of organic matter and minerals and macro-lamination, with lateral variation in properties due to compositional differences, both contribute to shale anisotropy of mechanical properties.
	
\begin{figure}
		\centering
		\includegraphics[width = 1.0\textwidth]{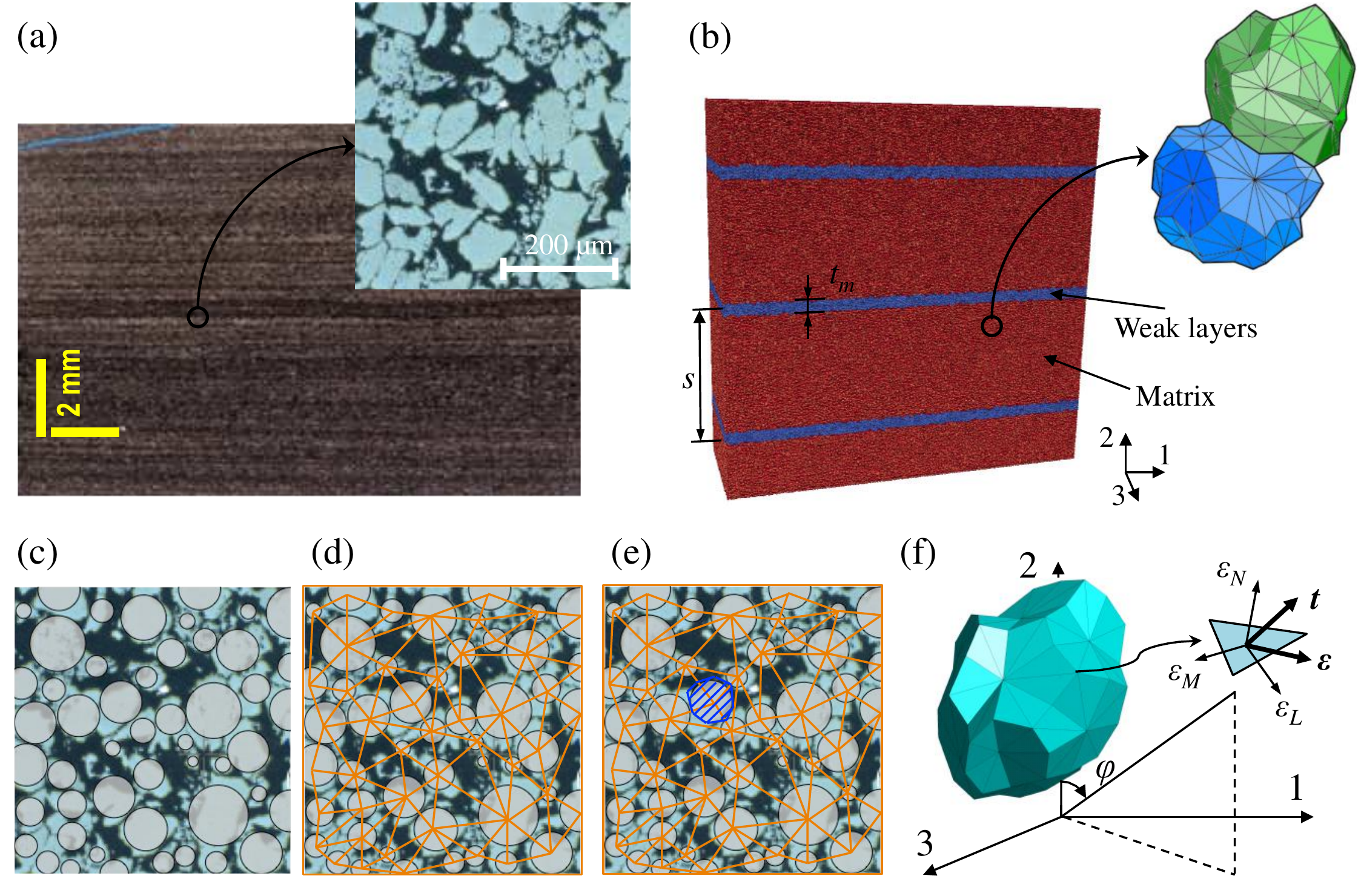}
		\caption{(a) Fine lamination in Barnett shale \cite{abouelresh2012lithofacies} and granular microstructure of Toarcianshale shale \cite{akono2016microscopic}. (b) A LDPM system representing the laminated structure model and a zoomed view of grain interaction. (c) Spherical particles placed at the center of shale grains for grain generation. (d) 2D representation of Delaunay tetrahedralization connecting spherical particles. (e) 2D representation of a polyhedral cell generated by a domain tessellation. (f) A 3D polyhedral cell representing a shale grain. }
		\label{fig:ShaleLDPM}
	\end{figure}

	Based on the composition and texture of shale described above, a shale sample at the mesoscopic level (scale 5 in Figure \ref{fig:MultiscaleShale}) can be represented by a laminated structure model with multiple layers of weakness embedded in a stiffer, stronger, and tougher matrix (Figure \ref{fig:ShaleLDPM}b), which is an analogy for the typical laminated shale as shown in Figure \ref{fig:ShaleLDPM}a. The matrix behavior outside the weak layers is dominated by the constituent components of shale, including organic matter, clay, quartz, and other minerals, which also exhibit anisotropic behavior. Hence, in addition to explicitly simulating the weak layers, a smeared representation of transverse isotropy in the matrix is also introduced. The smeared approach accounts for sources of anisotropy characterized by a length scale smaller than few micrometers. The transversely isotropy assumption is employed due to the fact that the sources of shale anisotropy at the microscopic level. Thus, the plane of isotropy can be considered to coincide with the plane of bedding/lamination. In the proposed formulation, the laminated structure is simulated by assigning different material properties for the layers of weakness and matrix. 
	
	The behavior of the weak layers as well as of the matrix is simulated by means of LDPM, which adopts a strategy to replicate the grain-scale heterogeneity of shale as depicted in Figure \ref{fig:ShaleLDPM}c-e. As the grains in shale tend to be densely packed and in direct contact with each other, a simplification is adopted to discretize the material domain into a granular lattice system without the isolation of cement bridges and grains. In this way, the contribution of cementing phase is embedded implicitly into the particle-scale constitutive laws controlling the interaction between skeletal grains. Similarly, the presence of microcracks, pores, and other inclusions at lower scales is also smeared out with grain interaction laws. 
	
	LDPM simulates the shale internal structure by considering only the coarser grains. An artificial supporting system with spherical particles placed at the center of shale grains (Figure \ref{fig:ShaleLDPM}c) is first generated following a strategy similar to that proposed by Cusatis et al. \cite{cusatis2011lattice1}. The size distribution of the spherical supports follows a probability density function (pdf) defined as
	\begin{equation}\label{eqn:psd}
		f(d) = \frac{qd_0^q}{[ 1-(d_0/d_a)^q ] d^{q+1}}
	\end{equation}
	where $d_0$ is the minimum particle size, $d_a$ is the maximum particle size, and $q$ is a material parameter. Note that pdf in Eq. \ref{eqn:psd} is associated with a sieve curve in the form:
	$
	F(d) = \left( d/d_a \right)^{n_F}
	$
	where $n_F = 3-q$ is the sieve curve exponent. The volume fraction of simulated particles can be calculated as $v_{a0} = [1-F(d_0)]v_a = [1-(d_0/d_a)^{n_F}]v_a$ where $v_a$ is the particle volume fraction, and the total volume of simulated particles is $V_{a0} = v_{a0}V$ with $V$ representing the specimen volume. Particle diameters $d_i$ are calculated by sampling the cumulative distribution function (cdf) associated with Eq. \ref{eqn:psd}: $d_i = d_0[1 - P_i(1-d^q_0/d^q_a)]^{-1/q}$, where $P_i$ is a sequence of random numbers between 0 and 1 generated by a random number generator. New particles are generated until the total volume of generated spherical particles, $\widetilde{V}_{a0} = \sum_{i}(\pi d^3_i/6)$, exceeds $V_{a0}$. The generated particles are randomly distributed across the specimen on vertices, edges, surface faces, and interior volume through a algorithm that avoids particles overlapping. 
	
	The next step is to finalize the construction of shale internal structure by defining the topology of the grains through a Delaunay tetrahedralization and a 3D tessellation. The Delaunay tetrahedralization discretizes the domain of interest by a 3D mesh of tetrahedra (Figure \ref{fig:ShaleLDPM}d) with the vertices coinciding with the given particle centers. The 3D domain tessellation, anchored to the Delaunay tetrahedralization, creates a system of polyhedral cells. Details on the adopted domain tessellation can be found in \cite{cusatis2011lattice1}. By collecting all facets associated with a particle (Figure \ref{fig:ShaleLDPM}e), one can obtain a polyhedral cell representing a cement-coated grain (Figure \ref{fig:ShaleLDPM}f). The grain size distribution of the simulated shale can be obtained by computing the volume of each polyhedral cell. It turns out that the statistical distribution of their volume-equivalent sphere diameters is similar to that of the supporting particles \cite{esna2015micro2}.  It is worth noting that the parameters, $d_0$, $d_a$, $n_F$, and $v_a$, must be calibrated through a trial-and-error procedure according to the measured grain size distribution of the selected rock, as reported in \cite{esna2015micro2}.

	\subsection{LDPM kinematics and equilibrium}
	In the LDPM formulation, adjacent cells interact through triangular facets where they are in contact. Rigid-body kinematics is adopted to describe the mesostructure deformation with which strain and stress vectors at the facet level can be derived. 
	
	According to this assumption, the facet strains, one normal component $\epsilon_N$ and two shear components $\epsilon_M$ and $\epsilon_L$, are defined through the relative displacement at the centroid of a given facet, which read
	\begin{equation}\label{ldpm-strain}
		\epsilon_N = \frac{\mathbf{n}^T \llbracket {\mathbf{u}_{C}} \rrbracket}{\ell}; \qquad \epsilon_M = \frac{\mathbf{m}^T \llbracket {\mathbf{u}_{C}} \rrbracket}{\ell}; \qquad \epsilon_L = \frac{\mathbf{l}^T \llbracket {\mathbf{u}_{C}} \rrbracket}{\ell}
	\end{equation}
	where $\llbracket {\mathbf{u}_{C}} \rrbracket$ is the displacement jump vector calculated by the displacements and rotations of the nodes adjacent to the selected facet, $\ell$ = tetrahedron edge associated with the facet, and $\mathbf{n}$, $\mathbf{m}$, and $\mathbf{l}$ are unit vectors defining a local system of reference attached to each facet. The facet strains can then be used to compute the LDPM facet stresses, $\boldsymbol{t}_c = t_N \mathbf{n} + t_M \mathbf{m} + t_L \mathbf{l}$, through the LDPM constitutive law as reported in the next section. 
	
	Furthermore, translational and rotational equilibrium equations of the particle P$_I$ is
	\begin{equation} \label{motion-1}
		\sum_{\mathcal{F}_I} A \mathbf{t}^{IJ} + V^I \mathbf{b}^0 = 0
	\end{equation}
	and	
	\begin{equation}\label{motion-2}
		\sum_{\mathcal{F}_I}  A \mathbf c^I \times \mathbf {t}^{IJ} = 0
	\end{equation}
	in which $\mathcal{F}_I$ is the group of facets surrounding node P$_I$ and associated with each node pair $(I,J)$; $A$ = facet area; $V^I$ is the particle volume; $\mathbf{b}^0$ is the body force vector; $\mathbf{c}^{I}$ = vector connecting nodes $P_I$ to the facet centroid, see Fig. \ref{TwoScaleAnalysis}c. 	
	
	\subsection{LDPM constitutive equations}\label{sec:Const}
	
	\subsubsection{Elastic behavior}
	In LDPM, the elastic behavior is formulated by linear relations between normal/shear stresses and the corresponding strains,
	\begin{equation}
		\label{eq1}
		t_N = E_N \epsilon_N; \quad t_M = E_T \epsilon_M; \quad t_L = E_T \epsilon_L
	\end{equation}
	where $E_N$ is the effective normal modulus; $E_T = \alpha E_N$ is the effective shear modulus, and $\alpha$ is the shear-normal coupling parameter. For isotropic materials, $E_N$ and $\alpha$ are constants and assumed to be material properties that can be identified from experimental data in the elastic regime. In order to model the transverse isotropy of shale, $E_N$ and $E_T$ are assumed to be functions of spatial orientation, i.e. $E_N = E_N(\varphi)$ and $E_T = E_T(\varphi)$, where \(\varphi\) is defined as the angle between the normal vector to LDPM facets and the normal vector to lamination planes as shown in Figure \ref{fig:ShaleLDPM}f. The following functions are assumed in this study
	\begin{equation}\label{eqn:ENTheat}
		E_N(\varphi) = \left( \frac{\sin^2 \varphi}{E_{N1}} + \frac{\cos^2 \varphi}{E_{N0}} \right)^{-1} 
		= E_{N0} \left( \beta _N \sin^2 \varphi + \cos^2 \varphi \right)^{-1}
	\end{equation}
	\begin{equation}\label{eqn:ETTheat}
		E_T(\varphi) = \left( \frac{\sin^2 \varphi}{E_{T1}} + \frac{\cos^2 \varphi}{E_{T0}} \right)^{-1} 
		= E_{T0} \left( \beta _T \sin^2 \varphi + \cos^2 \varphi \right)^{-1}
	\end{equation}	
	\begin{equation}
		\alpha(\varphi) = \frac{E_T(\varphi)}{E_N(\varphi)} = \alpha _0 \frac{\beta_N \sin^2 \varphi + \cos^2 \varphi}{\beta_T \sin^2 \varphi + \cos^2 \varphi}
	\end{equation}		
	where $E_{N0} = E_N(0\si{\degree})$, $E_{N1} = E_N(90\si{\degree})$, $E_{T0} = E_T(0\si{\degree})$, and $E_{T1} = E_T(90\si{\degree})$. In addition, \(\beta_N = E_{N0}/E_{N1}\) and \(\beta_T = E_{T0}/E_{T1}\) are the ratios of the elastic moduli at \(0\si{\degree}\) to the ones at \(90\si{\degree}\); \(\alpha_0 = \alpha(\ang{0}) = E_{T0}/E_{N0}\) is the shear-normal coupling parameter at $\ang{0}$. 	
	
	At the continuum level, the elastic behavior of transversely isotropic materials can be characterized by five independent elastic constants, $E, E', \nu, \nu'$, and $G'$. \(E\) and \(\nu\) are the Young's modulus and Poisson's ratio in the plane of transverse isotropy; \(E'\) and \(\nu'\) are the ones in the plane perpendicular to the isotropy plane; \(G'\) is the out-of-plane shear modulus. The relationship between the mesoscale LDPM formulation and parameters and the equivalent continuum elastic parameters can be obtained by exploiting similarity between LDPM and the kinematically constrained formulation of the microplane model \cite{bazant2000microplane}. One has
	\begin{equation}
		\label{eqn:microplane}
		E^*_{ijkl} = \frac{3}{2\pi} \int_{\Omega} \left( E_N N_{ij}N_{kl} + E_T M_{ij}M_{kl} +E_T L_{ij}L_{kl} \right) d\Omega
	\end{equation}
	where $E^*_{ijkl}$ is the equivalent continuum stiffness tensor, $N_{ij} = n_in_j$, $M_{ij} = (m_i n_j + m_j n_i)/2$, and $L_{ij} = (l_i n_j + l_j n_i)/2$, in which $n_i$, $m_i$, and $l_i$ are local Cartesian coordinate vectors on the generic microplane or LDPM facet with $n_i$ being normal. Integration is conducted over a unit hemisphere with surface \(\Omega\) representing all possible microplane or facet orientations. 
	
	In general, Eq. \ref{eqn:microplane} cannot be inverted to obtain the four LDPM parameters from the five equivalent continuum parameters, which are typically obtained from experimental tests. As a matter of fact, it can be shown that the proposed LDPM formulation only covers a limited range of thermodynamically consistent Poisson's ratios $\nu$, $\nu'$. The full range can be obtained in the microplane model, and similarly in LDPM, if the microplane/facet formulation is based on the spectral decomposition of the elastic tensor \cite{congrui2015microplane}. Such formulation, however, limits significantly the ability of the model to simulate mechanical phenomena associated with material heterogeneity. Discussion of this aspect for the case of isotropic materials can be found in \cite{cusatis2011lattice1}.
	
	The effect of weak layers on the elastic properties is considered by introducing the reduction factors $C_N$ and $C_T$ for effective normal and shear moduli, as follows:
	\begin{equation}
		\label{eqn:ECN}
		E_N^l(\varphi) = C_N E_N; \quad E_T^l(\varphi) = C_T E_T
	\end{equation}
	The reduction factors satisfy \( 0 < C_N, C_T \leq 1\). To simplify the proposed model, it is assumed that these two reduction factors have the same value, i.e. \(C_N = C_T\). As a result, the macroscopic stiffness matrix of the weaker layers is proportional to that of the matrix. 
	
	\subsubsection{Fracturing behavior}
\begin{figure}
		\centering
		\includegraphics[width = 1.0\textwidth]{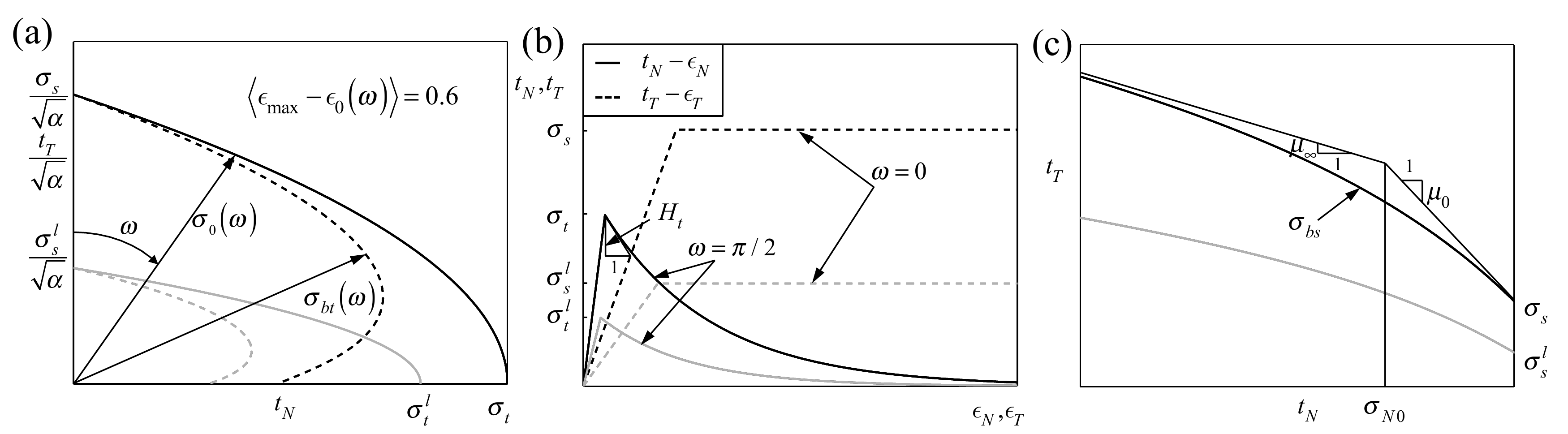}
		\caption{LDPM constitutive laws (gray lines represents the behaviors associated with weak layers): (a) shear strength as a function of normal tensile stresses; (b) typical stress versus strain curves at the LDPM facet levels; (c) shear strength as a function of normal compressive stresses. }
		\label{fig:Equations}
	\end{figure}

	The constitutive equations of inelastic fracturing in the LDPM formulation address the fracturing and cohesive behavior under tension and tension/shear for $\epsilon_N > 0$. To define the fracture and damage evolution, it is useful to first define the following effective stress \(t\) and effective strain \(\epsilon\): 
	\begin{equation}
		t = \sqrt{t_N^2 + (t_M^2 + t_L^2)/\alpha}; \quad \epsilon = \sqrt{\epsilon_N^2 + \alpha (\epsilon_M^2 + \epsilon_L^2)}
	\end{equation}
	The relationship between normal and shear stresses versus normal and shear strains can then be calculated in a way similar to simple damage models:
	\begin{equation}
		t_N = t \frac{\epsilon_N}{\epsilon}; \quad t_M = t \frac{\epsilon_M}{\epsilon}; \quad t_L = t \frac{\epsilon_L}{\epsilon}
	\end{equation}
	An internal variable $\omega$ characterizes the coupling between normal strain $\epsilon_N$ and total shear strain $\epsilon_T = \sqrt{\epsilon_M^2 + \epsilon_L^2}$ as $\tan (\omega) = \epsilon_N / \sqrt{\alpha} \epsilon_T$. The effective stress \(t\) is incrementally elastic ($\dot{t} = E_N \dot{\epsilon}$), and must satisfy the inequality \( 0 \leq t \leq \sigma_{bt} (\epsilon, \omega)\), where the strain-dependent boundary \(\sigma_{bt} (\epsilon, \omega)\) can be expressed as
	\begin{equation}\label{eqn:sigmabt}
		\sigma_{bt}(\epsilon,\omega) = \sigma_0(\omega) \text{exp} \left[-\frac{H_0(\omega)}{\sigma_0(\omega)} \langle \epsilon_{\text{max}} - \epsilon_0(\omega) \rangle \right]
	\end{equation}
	in which the bracket \(\langle \cdot \rangle\) are used in Macaulay sense: \(\langle x \rangle = \text{max}\{ x, 0 \}\). The function \(\sigma_0(\omega)\) is the strength limit for the effective stress and is defined as
	\begin{equation}
		\label{strengthlimit}
		\sigma_0(\omega) = \sigma_t \frac{-\sin \omega + \sqrt{\sin^2 (\omega) + 4\alpha \cos^2(\omega)/r_{st}^2}}{2\alpha \cos^2 (\omega)/r_{st}^2}
	\end{equation}
	where \(r_{st} = \sigma_s/\sigma_t\) is the ratio between the shear (cohesion) strength \(\sigma_s\) and the tensile strength \(\sigma_t\). 	After the maximum effective strain reaches its elastic limit \(\epsilon_0(\omega) = \sigma_0 / E_N\), the boundary \(\sigma_{bt}\) decays exponentially according to Eq. \ref{eqn:sigmabt}. Figure \ref{fig:Equations}a illustrates the virgin and damaged strength domain computed from the boundary $\sigma_0$ and $\sigma_{bt}$, respectively. For pure tensile stress ($\omega = \pi/2$), the tensile boundary represents strain softening with exponential decay, characterized by the microscale tensile strength $\sigma_t$ and the softening modulus $H_t$; for pure shear stress ($\omega = 0$), the effective stress boundary represents perfectly plastic behavior characterized by the microscale shear strength $\sigma_s$. The decay rate is governed by the power-law function $H_0(\omega) = H_t \left( {2\omega}/{\pi} \right)^{n_t}$, where the material parameter $n_t$ allows for a nonlinear transition. Fig. \ref{fig:Equations}b shows typical stress versus strain curves for $\omega = 0$ and $\omega = \pi/2$. 
	
	Characterized by microscale tensile and shear strengths, the evolution of the effective stress boundary is also assumed to be orientation-dependent to address damage anisotropy at the microscopic level. In this study, tensile and shear strengths are considered as functions of \(\varphi\) with the functional forms similar to the ones for the elastic moduli (Equations \ref{eqn:ENTheat} and \ref{eqn:ETTheat}),
	\begin{align}\label{eqn:s_st}
		\sigma_t(\varphi) &= \sigma_{t0}(\beta_t \sin^2 \varphi + \cos^2 \varphi)^{-1} \\
		\sigma_s(\varphi) &= \sigma_{s0}(\beta_s \sin^2 \varphi + \cos^2 \varphi)^{-1}
	\end{align}
	where $\sigma_{t0} = \sigma_t (\ang{0})$, $\sigma_{t1} = \sigma_t (\ang{90})$, $\sigma_{s0} = \sigma_s (\ang{0})$, and $\sigma_{s1}
	= \sigma_s (\ang{90})$; \(\beta_t = \sigma_{t0} / \sigma_{t1}\) and \(\beta_s = \sigma_{s0} / \sigma_{s1}\).
	
	Similar to modeling elastic behavior, the effect of lamination can be considered by introducing the reduction factors \(C_{Nt}\) and \(C_{Ts}\). The tensile and shear strengths associated with weak layers can then be written as
	\begin{equation}
		\sigma_t^l(\varphi) = C_{Nt} \sigma_t; \quad \sigma_s^l(\varphi) = C_{Ts} \sigma_s
	\end{equation} 
	in which \( 0 < C_{Nt}, C_{Ts} \leq 1\). As a result, the strength limit for the effective stress $\sigma_0 (\omega)$ is a function of the facet orientation \(\varphi\) and location in addition to $\omega$. Equation \ref{strengthlimit} is a parabola in $t_N - t_T$ space with the axis of symmetry along the \(t_N\)-axis. The parabola represents the envelop of the elastic limit of stress status $(t_N, t_T)$, varies with facet orientation, and shrinks for facets associated with weak layers (solid curves in Fig. \ref{fig:Equations}a). 
	
	To preserve the correct energy dissipation during microscale damage localization, the softening modulus in pure tension is expressed as \(H_t = 2 E_N / (\ell_t / \ell -1)\), where the characteristic length $\ell_t = 2 E_N G_t / \sigma_t^2$, \(G_t\) is the microscale fracture energy, and $\ell$ is the length of the tetrahedron edge associated to the current facet. Although a orientation-dependent relationship can be also applied to softening modulus \(H_0 (\omega)\), fracture energy \(G_t\), characteristic length \(l_t\), and \(n_t\) at the facet level, lacking any laboratory observation on the post-peak behavior of anisotropic shale in response to tensile loading makes it infeasible. Therefore, $\ell_t$ and $n_t$ are considered as constants in the current work. One can obtain the expressions of the microscale fracture energy from $E_N(\varphi)$ and $E_T(\varphi)$ functions, which read
	\begin{equation}
		G_t(\varphi) = \frac{l_t \sigma_t^2}{2 E_N} = \frac{l_t \sigma_{t0}^2 \left( \beta_t \sin^2 \varphi + \cos^2 \varphi \right)^{-2}}{2 E_{N0} \left( \beta_N \sin^2 \varphi + \cos^2 \varphi \right)^{-1} }
	\end{equation}
	for the matrix, and
	\begin{equation}
		G_t^l(\varphi) = \frac{C_{Nt}^2}{C_N} G_t
	\end{equation}
	for the weak layers. 
	
	\subsubsection{Frictional behavior}
	Due to frictional effects, the shear strength increases in the presence of compressive stresses. This phenomena can be simulated effectively through classical incremental plasticity. Incremental shear stresses are calculated as $\dot{t}_M = E_T (\dot{\epsilon}_M - \dot{\epsilon}^P_M)$ and $\dot{t}_L = E_T (\dot{\epsilon}_L - \dot{\epsilon}^P_L)$ where the plastic strain increments are assumed to follow the normality rule. The plastic potential can be expressed as \(\varphi = \sqrt{t_L^2 + t_M^2} - \sigma_{bs}(t_N) \) in which the shear boundary is formulated with the following frictional law: 
	\begin{equation}
		\sigma_{bs}(t_N) = \sigma_s + (\mu_0 - \mu_{\infty}) \sigma_{N0} - \mu_{\infty} t_N - (\mu_0 - \mu_{\infty}) \sigma_{N0} \exp(t_N / \sigma_{N0})
	\end{equation}
	where \(\mu_0\) and \(\mu_\infty\) are the initial and finial internal friction coefficients; \(\sigma_{N0}\) is the normal stress at which the internal friction coefficient transitions from \(\mu_0\) to \(\mu_{\infty}\). It can be seen from Fig. \ref{fig:Equations}c that in the presence of compressive stresses, the shear strength increases due to frictional effects. Since there are no available experimental data, we simply adopt the classical Coulomb linear frictional law with slop \(\mu_0\) by setting \(\sigma_{N0} = \infty\) or \(\mu_{\infty} = \mu_0\).
	
	The shear boundary governs the unconfined and low confinement macroscopic compression for low values of \(t_N\) and high confinement macroscopic compression for high values of \(t_N\). To address the damage anisotropy in the compression-shear regime, we allow the internal friction coefficient to vary in a way similar to the elastic moduli, i.e. 
	\begin{equation}
		\mu_0(\varphi) = \mu_{00} (\beta_{\mu} \sin^2 \varphi + \cos^2 \varphi)^{-1}
	\end{equation}
	\begin{equation}
		\mu_0^l(\varphi) = C_{\mu} \mu_0
	\end{equation}
	where $\mu_{00}$ is the initial internal friction coefficient at $\varphi = \ang{0}$; $\beta_{\mu}$ is the ratio of the friction coefficient at $\ang{0}$ to the one at $\ang{90}$; the reduction factor $C_{\mu}$ considers the effect of lamination on internal friction.  
	In addition, LDPM has the capability to simulate the triaxial compressive behavior at the macroscopic scale with a compressive boundary capturing pore collapse, material compaction, and rehardening. Since this is outside our current research scope, an elastic behavior is assumed for compression, i.e. \(t_N = E_N \epsilon_N\) for $\epsilon_N < 0$.
	
	To summarize, the LDPM material parameters governing shale elastic, fracturing, and frictional behaviors are listed in Table \ref{tab:LDPMpara}. It is also worth mentioning that the microscale constitutive relations discussed above incorporate the fine-scale processes at the interface between grains, which indirectly reflect the role of sub-resolution properties that are not explicitly modeled by LDPM. 
	
	\begin{table}
		\caption{LDPM material parameters for shale}
		\label{tab:LDPMpara}
		\centering
		\begin{tabular}{ll}
			\toprule
			Paramaters & Units\\
			\midrule
			Normal modulus \(E_{N0}\) & GPa \\
			Ratio of normal modulus \(\beta_N\) & \\
			Lamination reduction factor for normal modulus \(C_N\) & \\
			Shear modulus \(E_{T0}\) & GPa \\
			Ratio of shear modulus \(\beta_T\) & \\
			Lamination reduction factor for shear modulus \(C_T\) & \\
			Tensile strength \(\sigma_{t0}\) & MPa \\
			Ratio of tensile strength \(\beta_t\) & \\
			Lamination reduction factor for tensile strength \(C_{Nt}\) & \\	
			Shear strength \(\sigma_{s0}\) & MPa \\
			Ratio of tensile strength \(\beta_s\) & \\
			Lamination reduction factor for shear strength \(C_{Ts}\) & \\	
			Characteristic length \(l_t\) & mm \\
			Softening exponent \(n_t\) & \\
			Initial frictional coefficient \(\mu_{00}\) & \\
			Ratio of initial frictional coefficient \(\beta_{\mu}\) & \\
			Lamination reduction factor for initial friction \(C_{\mu}\) & \\
			\bottomrule
		\end{tabular}
	\end{table}		
	
	\section{Parametric study and calibration}
	The proposed model is implemented into the MARS software \cite{pelessoneMARS}, which is a structural analysis computer code with an object-oriented architecture that makes the implementation of new computational technologies very effective. Calibration is performed by comparing numerical simulation results with experimental data gathered from literature. The experiments considered hereinafter were conducted by J. Cho et al. \cite{cho2012deformation,kim_anisotropy_2012} on Boryeong shale. 
	
	As discussed above, the internal structure of shale is mimicked by a laminated granular lattice system governed by a set of geometric-related parameters at micro- and meso- scales. Observation and measurements of the materials' internal structure under study and an investigation of its grain size distribution are required to calibrate these parameters. Since they are not currently available in literature for the shale under consideration, reasonable assumptions are made based on microanalysis on other types of shale and mudstone \cite{akono2016microscopic, mokhtari2014tensile, vernik1992ultrasonic, slatt2011merging,eseme2006factors}. The spacing of lamination $s$ is assumed to be \SI{1}{mm}, and the thickness of weak layers $t_m$ about two times the typical grain size; the maximum, minimum, and mean grain sizes are assumed to be \SI{50}{\micro\meter}, \SI{15}{\micro\meter}, and \SI{30}{\micro\meter} respectively. The algorithm for the generation of the LDPM lattice system discussed in Section \ref{sec:Geom} is used to approximate the given granular structure. The resulting parameters $d_0 = \SI{15}{\micro\meter}$, $d_a = \SI{35}{\micro\meter}$, $n_F = 0.5$, and $v_a = 0.55$ generate a grain size distribution with a maximum, minimum, and mean diameter of \SI{51}{\micro\meter}, \SI{13}{\micro\meter}, and \SI{30}{\micro\meter}, respectively. 
	
\begin{figure}
		\centering
		\includegraphics[width = 0.6\textwidth]{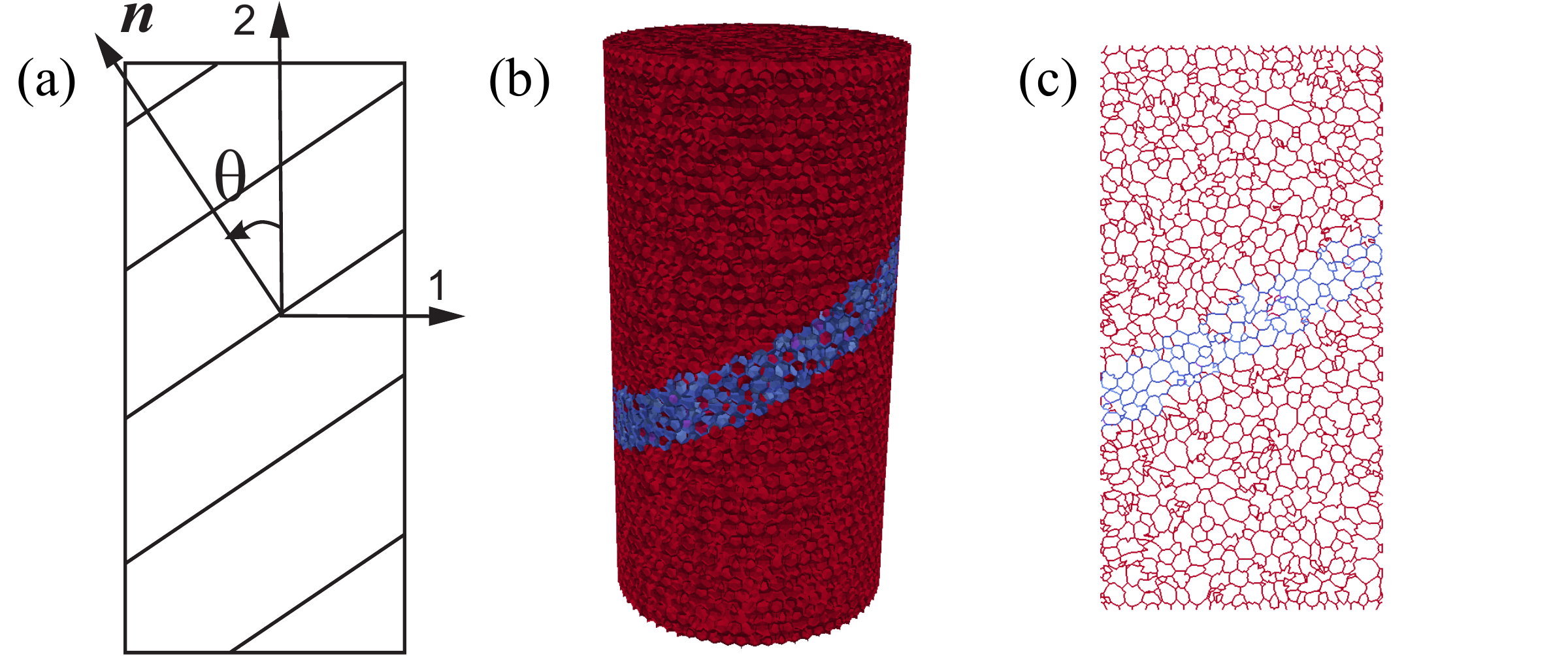}
		\caption{Cylindrical specimens in the uniaxial compression tests: (a) anisotropy angle $\theta$ definition; (b) a 3D granular lattice system representing laminated shale specimens; (c) a slice view in 1-2 plane of the polyhedral cell assemblage.}
		\label{fig:UCCANISO}
	\end{figure}

\begin{figure}
		\centering
		\includegraphics[width = 0.7\textwidth]{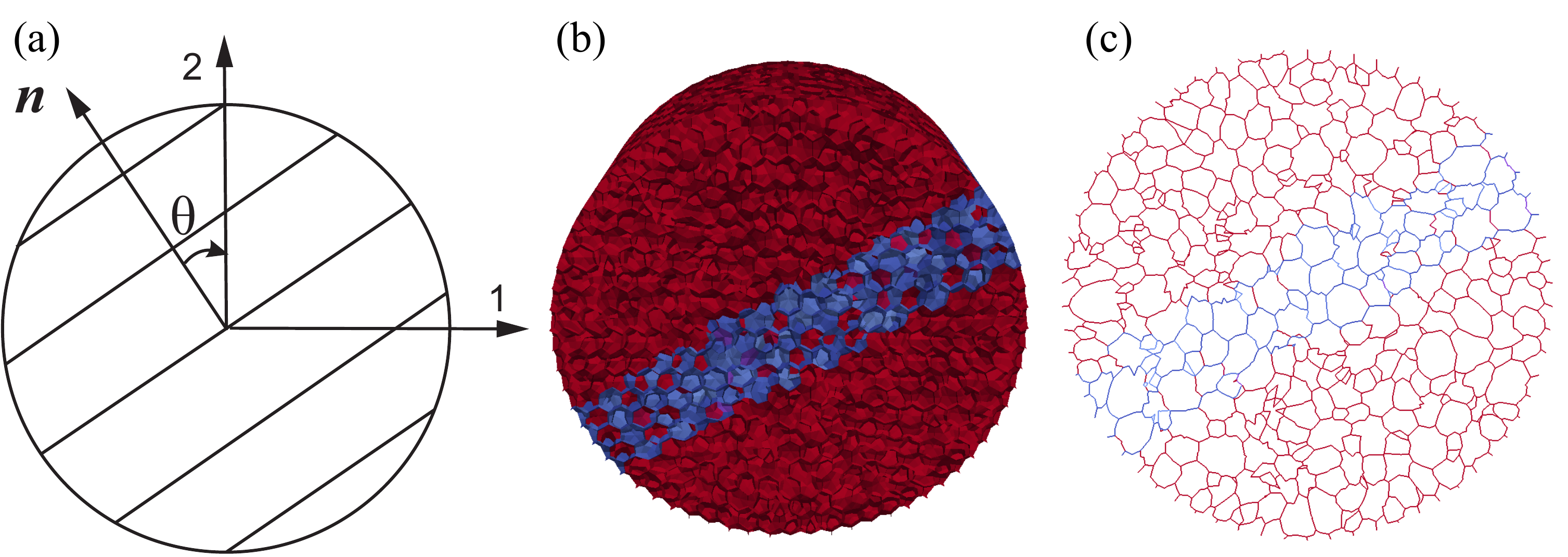}
		\caption{Disk specimens in the Brazilian tensile tests: (a) 2D rotation of the anisotropy angle $\theta$; (b) a 3D granular lattice system representing laminated shale specimens; (c) a slice view in 1-2 plane of the polyhedral cell assemblage. }
		\label{fig:BRZANISO}
	\end{figure}

	Cylindrical specimens with the diameter of \SI{0.5}{mm} and height to diameter ratio of 2 are generated for the simulation of uniaxial tests; disc specimens with the thickness and radius of \SI{0.25}{mm} are generated for the simulation of Brazilian tests. Each type consists of seven specimens with anisotropy angles of \ang{0}, \ang{15}, \ang{30}, \ang{45}, \ang{60}, \ang{75}, and \ang{90} respectively. The anisotropy angle for simulated specimens is defined as the angle between the loading direction and the normal vector to the plane of lamination, as shown in Figures \ref{fig:UCCANISO}a and \ref{fig:BRZANISO}a. Since the simulation of the actual sample size would lead to tens of millions of grains and to excessive computational cost, the simulated specimens for the calibration study are much smaller than the actual ones. As a result, the generated specimens consist of only one layer of weakness, which, however, are typical of the whole mixture on average, and are considered to be able to preserve the main failure mechanisms.
	
	In the proposed model, the macroscopic properties of material are determined by the microscale parameters governing the facet constitutive law listed in Table \ref{tab:LDPMpara}. Since direct measurement of these parameters is not available at the moment, a computationally intensive calibration task is required. Preliminarily, a parametric study is carried out to investigate the relationship between mesoscale parameters and macroscopic properties, which serves as a guide for the general procedure of parameter selection for other types of shale and layered rocks. 	
	
	\subsection{Elastic analysis} \label{sec:elasticanalysis}
	The five elastic constants for Boyeong shale were measured by J. Cho et al.\cite{cho2012deformation} through uniaxial compression tests on cylindrical specimens, which are simulated in this section. The specimens are loaded through two steel platens of their top and bottom ends. The values of simulated Young's modulus were calculated by $\sigma_{33}/\varepsilon_{33}$, where $\sigma_{33}$ is the macroscopic uniaxial stress, and $\varepsilon_{33}$ is the macroscopic uniaxial strain. In this paper, the macroscopic uniaxial stress $\sigma_{33}$ is approximated by a nominal stress, i.e. $\sigma_{33} = P/A$, where $P$ is the load applied on the specimen, and $A$ is the area of its cross section; the macroscopic uniaxial strain $\varepsilon_{33}$ is approximated by a nominal strain, i.e. $\varepsilon_{33} = \Delta L / L$, where $\Delta L$ is the specimen length change, and $L$ is the original length of the specimen.
	
	The calibration is completed by optimizing the LDPM parameters related to elastic behavior, i.e. $E_{N0}$, $\beta_{N}$, $C_N$, $E_{T0}$, $\beta_T$, and $C_T$, through the best fitting of the experimental data with the simulation results. In the case of $C_N = C_T = 1$, the remaining four LDPM parameters can be approximated by solving a continuous optimization problem which minimizes the global discrepancy between the norm of shale macroscopic stiffness matrix and the corresponding quantity given by Equation \ref{eqn:microplane}. The optimization problem takes the following form:
	\begin{equation}
		\begin{aligned}
			&\underset{\{E_{N0}, \beta_{N}, E_{T0}, \beta_T\} }{\operatorname{minimize}}& & \left\Vert E_{ijkl} - E^*_{ijkl}\right\Vert \\			
			&\operatorname{subject \quad to}
			& &  E_{N0}, \beta_{N}, E_{T0}, \beta_T \geq 0\\
		\end{aligned}
	\end{equation}
	where $E_{ijkl}$ is the macroscopic stiffness tensor of shale obtained via laboratory measurements, and $E^*_{ijkl}$ are obtained by solving the surface integral in Eq. \ref{eqn:microplane} numerically with 37 microplanes of different orientations \cite{bavzant1986efficient}. The optimization problem is solved with the help of the commercial software Mathematica \cite{math}. The optimal solution $\{E_{N0}^*, \beta_{N}^*, E_{T0}^*, \beta_T^*\}$ is given in the last row of Table \ref{tab:ElastLDPM}. They are treated as a set of input parameters and fed into the proposed micromechanical model. The elastic moduli obtained through the simulations of uniaxial compression tests are compared with the experimental data to verify the calculation. 
	
\begin{figure}
		\centering
		\includegraphics[width = 0.6\textwidth]{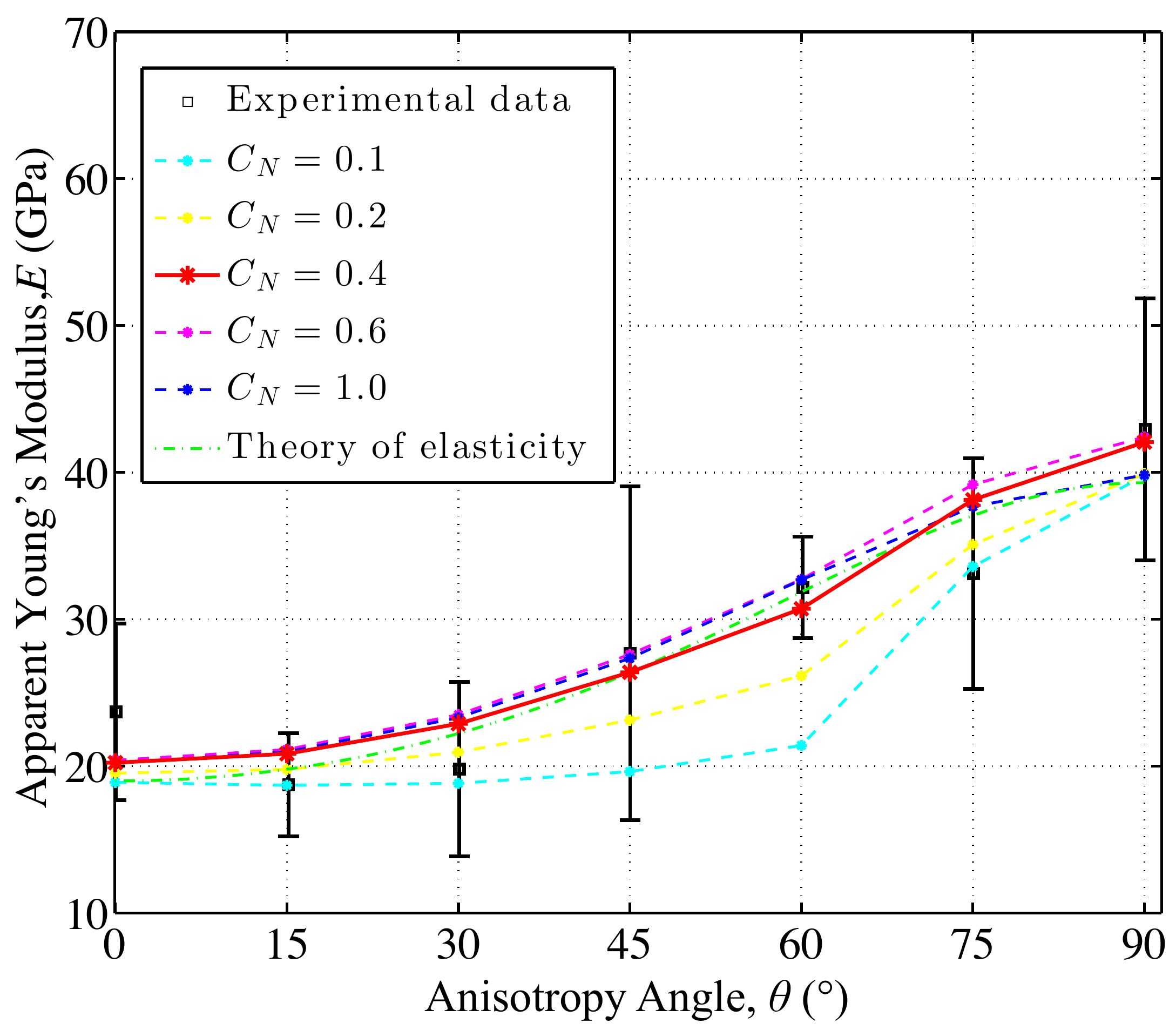}
		\caption{Comparison of apparent Young's modulus from experiments and simulations as a function of anisotropy angle given different values of $C_N$. }
		\label{fig:Young}
	\end{figure}

	Figure \ref{fig:Young} compares the variations of Young's modulus with anisotropy angle obtained from experiments and simulations. The error bar shows the experimental data, while the blue dash curve represents the numerical results with $C_N = C_T = 1$. The green dash-dot curve shows the variation of Young's modulus predicted by the theory of elasticity. It can be found that the numerical results with $C_N = C_T = 1$ are consistent with the ones predicted by the theory of elasticity. 
	
	\begin{table}
		\caption{Calibrated LDPM elastic parameters given different value of $C_N \& C_T$}
		\label{tab:ElastLDPM}
		\centering
		\begin{tabular}{llllll}
			\toprule
			$C_N = C_T$ & $E_{N0}$ (GPa) &$\beta_N$	& $E_{T0}$ (GPa) & $\beta_T$&\\
			\midrule	
			0.1	& 44.4		& 0.54		& 8.0	& 0.23		\\
			0.2	& 94.1		& 0.23		& 19.2	& 0.56		\\
			0.4	& 94.0		& 0.24		& 4.0	& 0.13		\\
			0.6	& 100.3		& 0.23		& 3.7	& 0.10		\\
			1	& 13.5		& 0.24		& 4.9	& 0.22		\\
			\bottomrule
		\end{tabular}
	\end{table}		
	
	The effect of $C_N$ and $C_T$ on simulated Young's modulus is also investigated herein. Given different values of $C_{N}$ and $C_{T}$ ($C_N = C_T = $ 0.1, 0.2, 0.4, and 0.6, respectively), the remaining parameters are calibrated through the best fitting of the measured Young's moduli of the specimens with $\theta = \ang{0}$ and $\theta = \ang{90}$ via a procedure similar to the one discussed above. The best fitting in each case of $C_N$ and $C_T$ is achieved using the input parameters listed in Table \ref{tab:ElastLDPM}. The variations of simulated Young's modulus given different values of $C_N$ and $C_T$ are represented by curves with different colors as illustrated in Figure \ref{fig:Young}, and compared with experimental data as well as the one predicted by the theory. It is worthing noting that although each layer is approximated by the transversely isotropic media, the composite response deviates from the theoretical prediction based on the transverse isotropy assumption as $C_N = C_T \neq 1$. The optimized value of $C_N$ is chosen such that the curve of the variation of Young's modulus from the simulations is in best agreement with the experimental data. One can find from Figure \ref{fig:Young} that the red solid curve with $C_N = C_T = 0.4$ has the best fit to the experimental data, and therefore determine the calibrated value to be 0.4. The corresponding LDPM parameters are listed in the 3rd row of Table \ref{tab:ElastLDPM}. 
	
	\subsection{Brazilian and uniaxial compression tests}
	The next step of the calibration process involves the parameters governing the anisotropic response to tensile and compressive loading up to failure. Brazilian and uniaxial compression tests on Boryeong shale which demonstrates significant anisotropy in the measured Brazilian tensile strength (BTS) and uniaxial compressive strength (UCS) are considered herein. In this study, the values of BTS are calculated based on the anisotropic solutions suggested by Claesson \cite{claesson_brazilian_2002}, while the values of UCS are approximated by the peaks of the nominal uniaxial stress $\sigma_{33}$. In the proposed model, the quantities of BTS and UCS and their variations with anisotropy angles are mostly governed by the normal and shear strengths as well as the lamination reduction factors. The parametric study involving the effects of $\sigma_t/\sigma_s$, $C_{Nt}$, $C_{Ts}$, and $C_{\mu}$ is presented in the following sections, which provides a fundamental understanding of how the LDPM parameters at the meso- and micro- scales control mechanical strength at the macroscopic level. 
	
	\subsubsection{Effect of $\sigma_s/\sigma_t$}\label{sec:STratio}
\begin{figure}
		\centering
		\includegraphics[width = 1\textwidth]{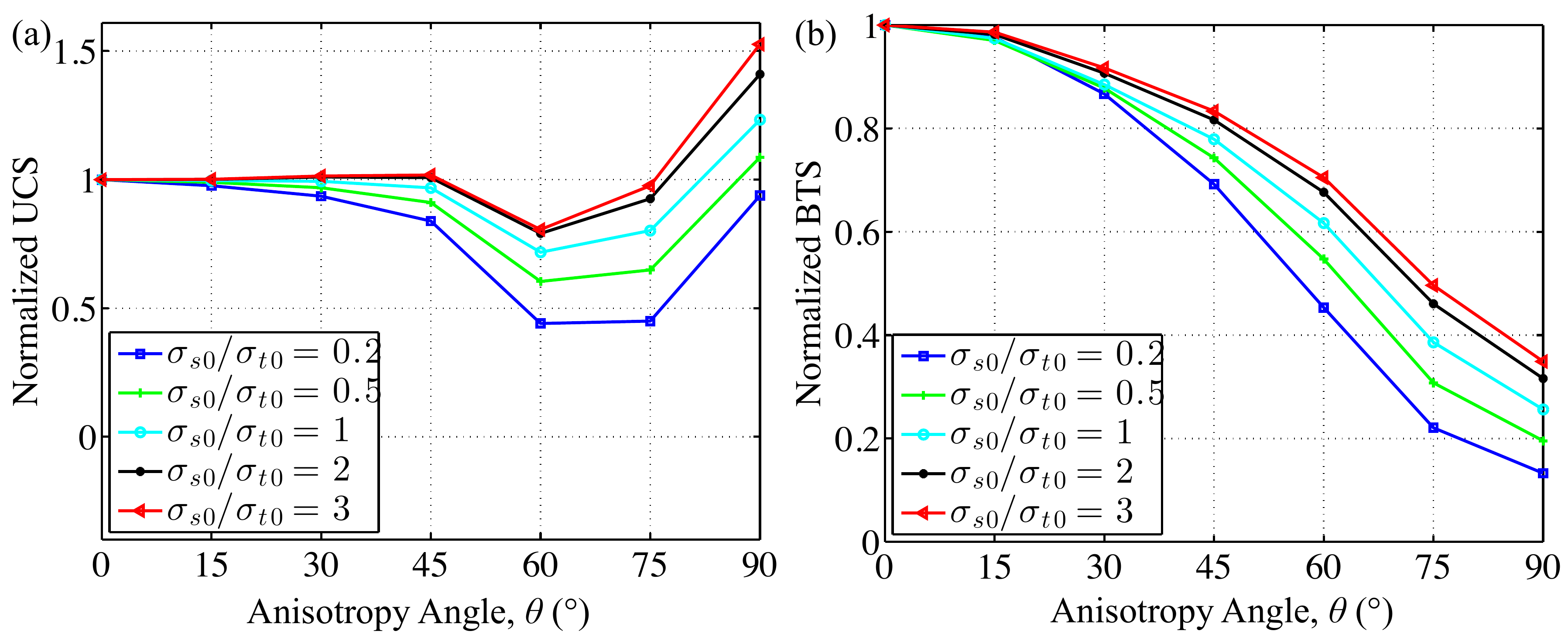}
		\caption{Effect of $\sigma_s/\sigma_t$: (a) normalized UCS and (b) normalized BTS.}
		\label{fig:STratio}
	\end{figure}

	The effect of ratio between cohesion and normal strength, i.e. $\sigma_s/\sigma_t$, is investigated here. To simplify the analysis, the parameters $\beta_t$ and $\beta_s$ are set to the same as $\beta_N$ and $\beta_T$ respectively. In this case, the effect can be explored by adjusting $\sigma_{t0}$ while keeping $\sigma_{s0}$ fixed. Brazilian and uniaxial compression tests are simulated with the ratio $\sigma_{s0}/\sigma_{t0}$ of 0.2, 0.5, 1, 2, and 3 respectively. The other parameters are kept the same as Table \ref{tab:LDPMvalue} shows. The variations of normalized UCS and BTS against different values of $\sigma_{s0}/\sigma_{t0}$ are shown in Figure \ref{fig:STratio}. In Table \ref{tab:STratio}, values of the anisotropy degree for UCS and BTS, the ratio of UCS at \ang{90} to the one at \ang{0}, and the ratio of UCS to BTS at \ang{0} given different values of $\sigma_{s0}/\sigma_{t0}$ are presented. They are labeled as UCS$_\text{MAX}/\text{UCS}_\text{MIN}$, BTS$_\text{MAX}/\text{BTS}_\text{MIN}$, UCS(\ang{90})$/$UCS(\ang{0}), UCS(\ang{0})$/$BTS(\ang{0}), respectively. 
	
	The value of $\sigma_{s0}/\sigma_{t0}$ has significant effect on the ratio of UCS at \ang{90} to the one at \ang{0}, i.e. UCS(\ang{90})$/$UCS(\ang{0}). As one can see from Figure \ref{fig:STratio}a and the 4th column of Table \ref{tab:STratio}, as the ratio of $\sigma_{s0}$ to $\sigma_{t0}$ increases, the value of UCS(\ang{90})$/$UCS(\ang{0}) also increases. Khanlari \cite{khanlari_evaluation_2014} mentioned that $\lambda$-shaped failure mode occurs at $\theta = \ang{0}$. In this failure mode, a singular tensile splitting failure type initiates along the specimen axis, which bifurcates into two shear planes across lamination. Therefore, the microscale matrix tensile strength dominates the uniaxial compressive strength at \ang{0}. As a result of a larger ratio of $\sigma_{s0}/\sigma_{t0}$,  a relatively small matrix tensile strength leads to a decrease of UCS(\ang{0}), which contributes to the gain of UCS(\ang{90})$/$UCS(\ang{0}). For simulated specimens with $\theta = \ang{90}$, the failure modes are different given different values of $\sigma_{s0}/\sigma_{t0}$. Given a relative large $\sigma_{s0}$ and a relative small $\sigma_{t0}$, i.e. a large ratio of $\sigma_{s0}/\sigma_{t0}$, tensile splitting along lamination is the dominant failure mode. For a small ratio of $\sigma_{s0}/\sigma_{t0}$, the shear failure mode is dominant. Although an increase of $\sigma_{s0}/\sigma_{t0}$ also decreases UCS(\ang{90}), the reduction effect is less significant than its effect on UCS(\ang{0}). As a result, the larger the value of $\sigma_{s0}/\sigma_{t0}$, the greater the ratio of UCS at \ang{90} to the one at \ang{0}. 
	
	For Brazilian tests, a relatively small tensile strength at microscale associated with a large value of $\sigma_{s0}/\sigma_{t0}$ decreases both BTS$_\text{MAX}$ and BTS$_\text{MIN}$, since tensile splitting failure is dominant for specimens with anisotropy angles of both $\ang{0}$ and $\ang{90}$. A decreases on BTS$_\text{MAX}/\text{BTS}_\text{MIN}$ as illustrated in the Figure \ref{fig:STratio}b and the 3rd column in Table \ref{tab:STratio} can be explained by the fact that its effect on failure along the lamination is less sensitive than that on failure along the matrix. 
	
	\begin{table}
		\caption{Effect of $\sigma_s/\sigma_t$ on UCS and BTS.}
		\label{tab:STratio}
		\centering
		\begin{tabular}{cccccc}
			\toprule
			$\sigma_{s0}/\sigma_{t0}$ & UCS$_\text{MAX}/\text{UCS}_\text{MIN}$ &BTS$_\text{MAX}/\text{BTS}_\text{MIN}$	& UCS(\ang{90})$/$UCS(\ang{0}) & \\
			\midrule	
			0.2	& 2.27	& 7.53	& 0.94\\
			0.5 & 1.80	& 5.11	& 1.09\\
			1	& 1.73	& 3.90	& 1.23\\
			2	& 1.78	& 3.16	& 1.41\\
			3	& 1.89	& 2.87	& 1.53\\
			\bottomrule
		\end{tabular}
	\end{table}		
	
	\subsubsection{Effect of $C_{Nt}$}\label{sec:CNT}
\begin{figure}
		\centering
		\includegraphics[width = 1\textwidth]{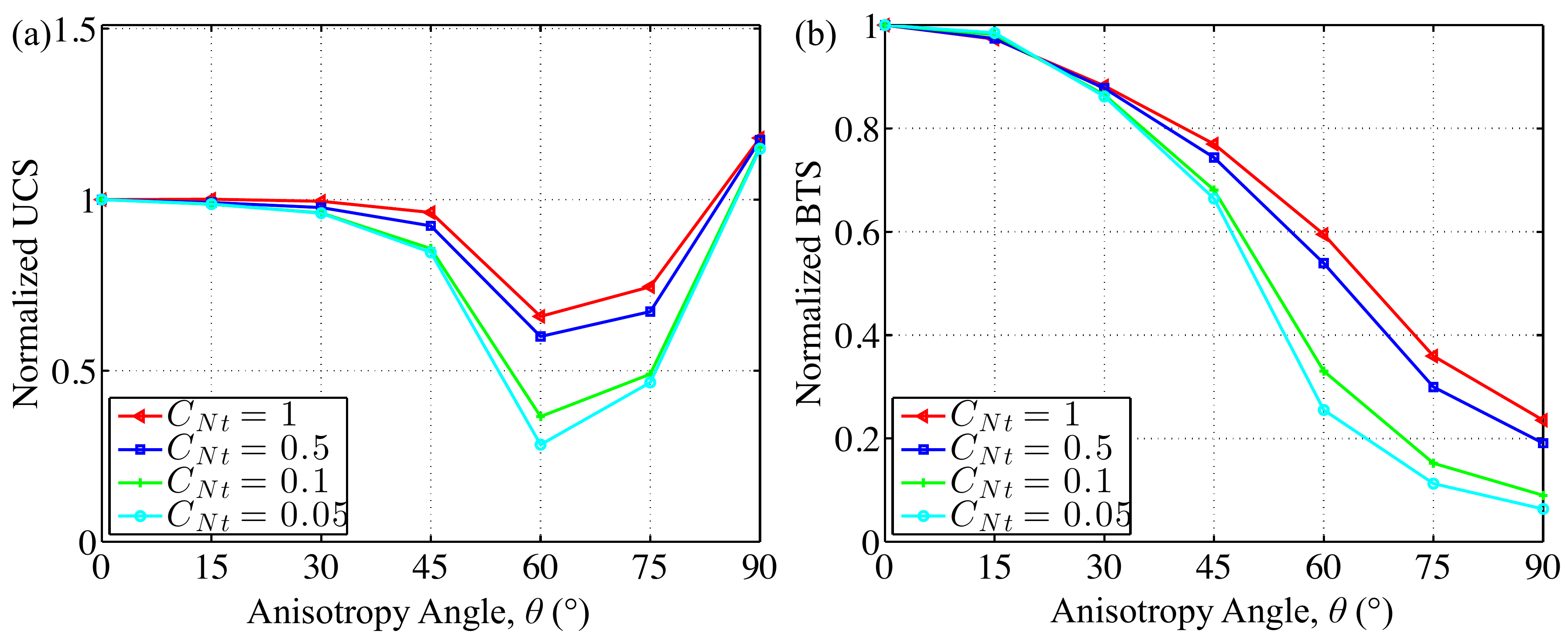}
		\caption{Effect of $C_{Nt}$: (a) normalized UCS and (b) normalized BTS.}
		\label{fig:CNT}
	\end{figure}

	The lamination reduction factor for tensile strength $C_{Nt}$ reduces the microscale tensile strength associated with weak layers, i.e. $\sigma_t^l$. Different values of $C_{Nt}$ are assigned to  weak layers, while other parameters are kept constant to investigate its effect on macroscopic BTS and UCS. The simulation results are illustrated in Figure \ref{fig:CNT} and listed in Table \ref{tab:CNT}.
	
	As one can observe from Figure \ref{fig:CNT}, the value of $C_{Nt}$ does not change the general trend of the normalized UCS and BTS variations. Especially, the influence of $C_{Nt}$ on the value of UCS(\ang{90})$/$UCS(\ang{0}) and the anisotropy angle at which the minimum UCS occurs can be neglected. However, as $C_{Nt}$ decreases, the anisotropy degree for UCS increases significantly, as illustrated in Figure \ref{fig:CNT}a and the 2nd column of Table \ref{tab:CNT}. Note that the change of $C_{Nt}$ dose not have significant influence on the value of UCS$_{\text{MAX}}$, because UCS$_{\text{MAX}}$ is mostly governed by failure along matrix, and thus less sensitive to $C_{Nt}$. On the contrary, a decrease of $C_{Nt}$ is closely related to the reduction of UCS$_{\text{MIN}}$ which typically occurs for $\ang{45} \leq \theta \leq \ang{75}$.  Although sliding failure along lamination is the dominant failure mode for $\ang{45} \leq \theta \leq \ang{75}$ at the macroscopic level, tensile failure also occurs along lamination following shear failure at microscale. Lisjak \cite{lisjak_continuumdiscontinuum_2014} and Liu \cite{liu2013numerical} illustrated numerically that both tensile cracking and shear cracking occur at the angles between \ang{45} and \ang{75}, and gradually develop into macro-cracks along lamination. Therefore, a lower microscale tensile strength related to weak layers reduces the load bearing capacity of specimens at $\ang{45} \leq \theta \leq \ang{75}$ under uniaxial compression, and finally leads to a decrease of the anisotropy degree for UCS at the macroscopic level. 
	
	Because tensile splitting failure along lamination occurs at \ang{90} for Brazilian test, it is obvious that tensile failure develops under a lower load level given a lower microscale tensile strength of weak layers. This explains that the anisotropy degree for BTS increases as $C_{Nt}$ increases. 
	
	\begin{table}
		\caption{Effect of $C_{Nt}$ on UCS and BTS.}
		\label{tab:CNT}
		\centering
		\begin{tabular}{cccccc}
			\toprule
			$C_{Nt}$ & UCS$_\text{MAX}/\text{UCS}_\text{MIN}$ &BTS$_\text{MAX}/\text{BTS}_\text{MIN}$	& UCS(\ang{90})$/$UCS(\ang{0}) & \\
			\midrule	
			1	& 1.79	& 4.25	& 1.18\\
			0.5 & 1.96	& 5.24	& 1.17\\
			0.1	& 3.15	& 11.11	& 1.15\\
			0.05& 4.04	& 15.71	& 1.15\\
			\bottomrule
		\end{tabular}
	\end{table}

	\subsubsection{Effect of \(C_{Ts}\)}\label{sec:CTS}
\begin{figure}
		\centering
		\includegraphics[width = 1\textwidth]{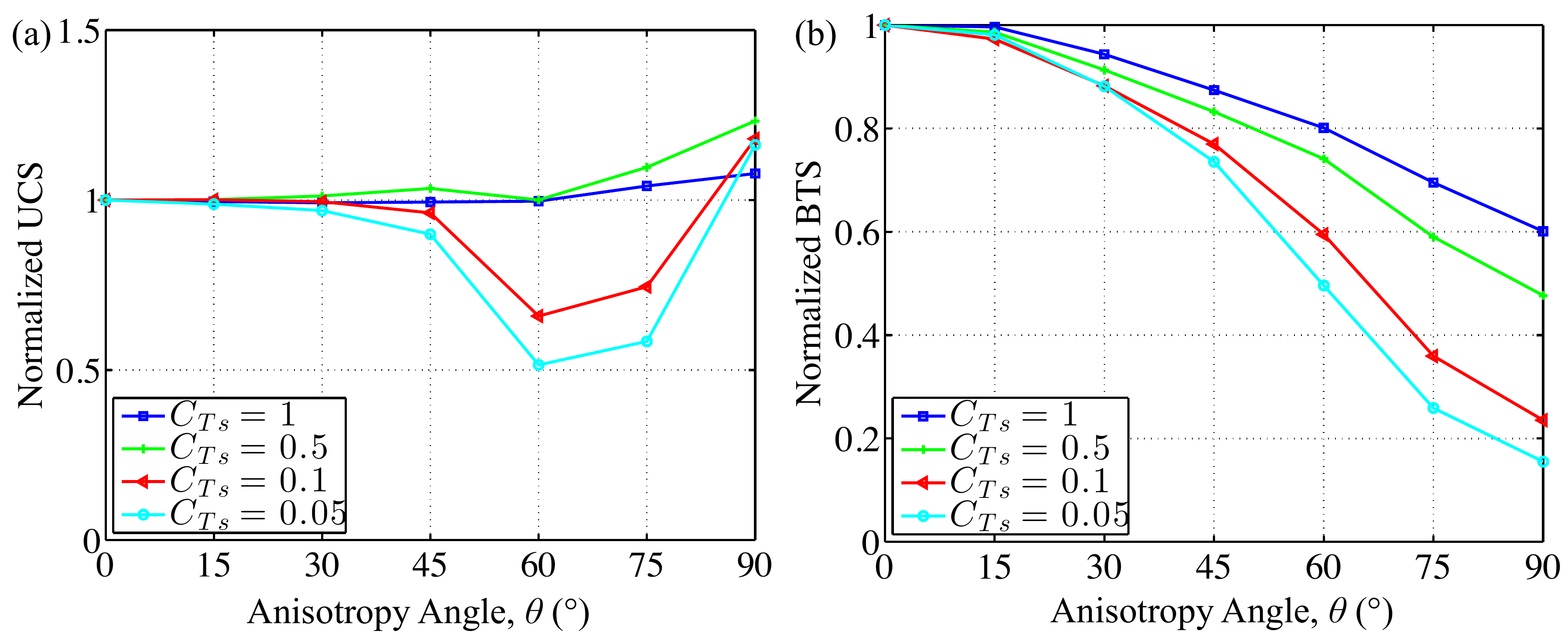}
		\caption{Effect of $C_{Ts}$: (a) normalized UCS and (b) normalized BTS.}
		\label{fig:CTS}
	\end{figure}

	The lamination reduction factor for shear strength $C_{Ts}$ is a key factor controlling the shear strength, $\sigma_s^l$ of weak layers. To investigate the effect of $C_{Ts}$, different values of $C_{Ts}$ ($C_{Ts} = $1, 0.5, 0.1, and 0.05 respectively) are assigned to weak layers, while the other parameters are kept constant. The simulation results are shown in Figure \ref{fig:CTS} and Table \ref{tab:CTS}. 
	
	Similar to the effect of $C_{Nt}$, as $C_{Ts}$ decreases, the anisotropy degrees for both UCS and BTS increase, as one can conclude from the 2nd and 3rd columns in Table \ref{tab:CTS} as well as Figure \ref{fig:CTS}. Due to lower shear strength of the weak layers, shear failure along lamination becomes easier to develop for specimens at $\ang{45} \leq \theta \leq \ang{75}$ under uniaxial compression. Particularly, one can observe from Figure \ref{fig:CTS}a that when $C_{Ts} = 1$, i.e. no reduction effect for shear strength associated with weak layers, the simulated compressive strength does not feature the classical strength minimum for jointed rocks. 
	
	Although the reduction of microscale shear strength associated with weak layers decreases the load bearing capacity of specimens at all angles in Brazilian tests, its effect is more significant for specimens with steeper angles of lamination leading to different failure modes for specimens at different anisotropy angles. For $\theta \leq \ang{30}$, the crack propagates mostly along the loaded diameter while for $\theta \geq \ang{30}$, the rock begins to fail along the lamination \cite{cho2012deformation,vervoort2014failure}. Fractures along lamination are observed for $\theta = \ang{90}$. This explains the phenomenon that the smaller the value of $C_{Ts}$, the greater the anisotropy degree for BTS. 
	
	\begin{table}
		\caption{Effect of $C_{Ts}$ on UCS and BTS.}
		\label{tab:CTS}
		\centering
		\begin{tabular}{cccccc}
			\toprule
			$C_{Ts}$ & UCS$_\text{MAX}/\text{UCS}_\text{MIN}$ &BTS$_\text{MAX}/\text{BTS}_\text{MIN}$	& UCS(\ang{90})$/$UCS(\ang{0}) & \\
			\midrule	
			1	& 1.09	& 1.66	& 1.08\\
			0.5 & 1.23	& 2.10	& 1.23\\
			0.1	& 1.79	& 4.25	& 1.18\\
			0.05& 2.25	& 6.44	& 1.16\\
			\bottomrule
		\end{tabular}
	\end{table}			
	
	\subsubsection{Effect of \(C_{\mu}\)}\label{sec:CMU}
\begin{figure}
		\centering
		\includegraphics[width = 1\textwidth]{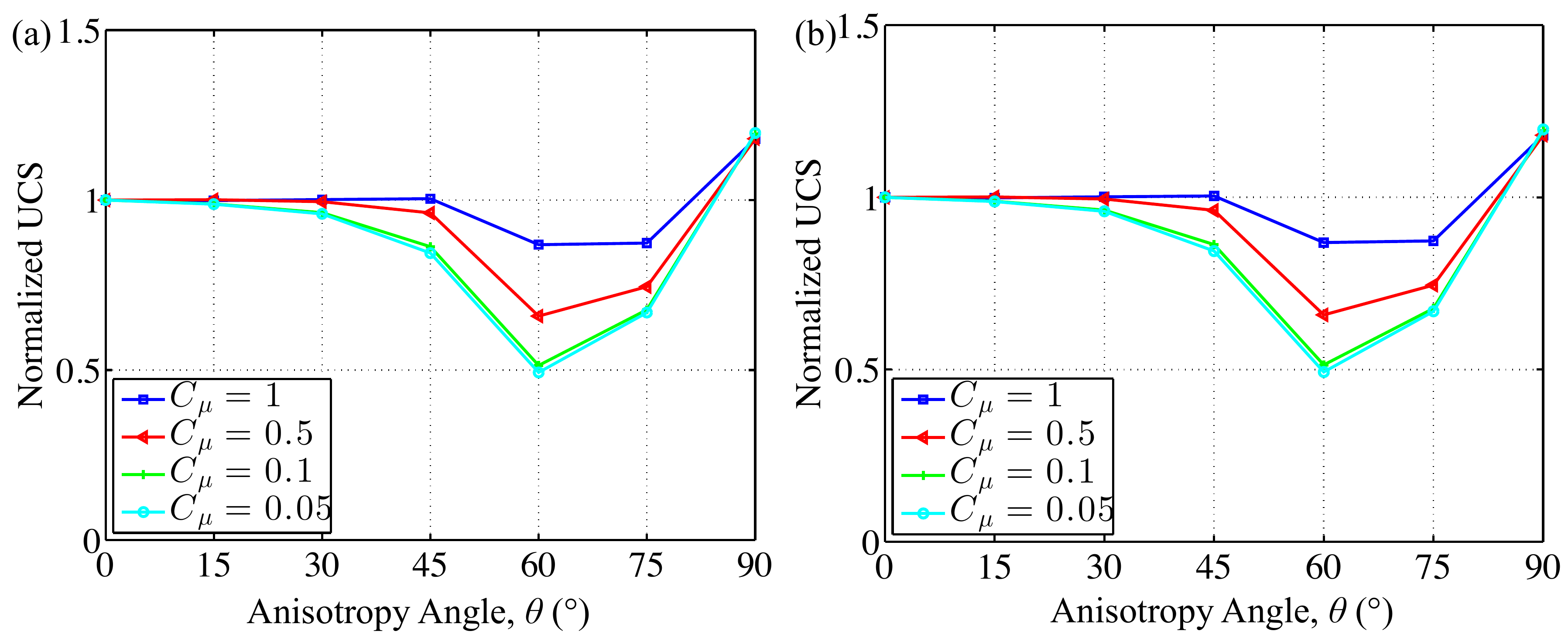}
		\caption{Effect of $C_{\mu}$: (a) normalized UCS and (b) normalized BTS.}
		\label{fig:CMU}
	\end{figure}

	The results of normalized UCS and BTS as a function of anisotropy angle with different lamination reduction factors for the internal friction $C_{\mu}$ are compared in Figure \ref{fig:CMU}. $C_{\mu}$ is another key factor controlling the shear boundary envelop associated with weak layers. The smaller the value of $C_{\mu}$, the smaller the friction angle, and thus, the smaller the slope of the microscale shear failure envelop related to weak layers. 
	
	As illustrated in Figure \ref{fig:CMU}a, the reduction of friction coefficient for weak layers significantly increases the anisotropy degree for UCS. It is presented in the 2nd column of Table \ref{tab:Cmu} that the anisotropy degree increases from \numrange{1.35}{2.43}, as the value of $C_{\mu}$ reduces from \numrange{1}{0.05}. Similar effect of $C_{\mu}$ can be noted on the anisotropy degree for BTS. However, it is less significant compared to its effect on the anisotropy degree for UCS. Note that when the value of $C_{\mu}$ is smaller enough, the effect of reducing anisotropy degrees becomes negligible as the shear failure envelop related to weak layers tends to become flat. 
	\begin{table}
		\caption{Effect of $C_{\mu}$ on UCS and BTS.}
		\label{tab:Cmu}
		\centering
		\begin{tabular}{cccccc}
			\toprule
			$C_{\mu}$ & UCS$_\text{MAX}/\text{UCS}_\text{MIN}$ &BTS$_\text{MAX}/\text{BTS}_\text{MIN}$	& UCS(\ang{90})$/$UCS(\ang{0}) & \\
			\midrule	
			1	& 1.35	& 3.55	& 1.18\\
			0.5 & 1.79	& 4.25	& 1.18\\
			0.1	& 2.32	& 4.93	& 1.19\\
			0.05& 2.43	& 4.98	& 1.20\\
			\bottomrule
		\end{tabular}
	\end{table}			
	
	\subsection{Calibrated results}\label{sec:Calib}
	Based on the results of the parametric study, the model calibration is performed, following a trial-and-error procedure similar to the one in section \ref{sec:elasticanalysis}. The microscale parameters for simulating uniaxial compression and Brazilian tests are calibrated by comparing the numerical results of UCS and BTS with the corresponding experimental data from \cite{cho2012deformation}. The finalized model parameters are reported in Table \ref{tab:LDPMvalue}. 
	
\begin{figure}
		\centering
		\includegraphics[width = 1.0\textwidth]{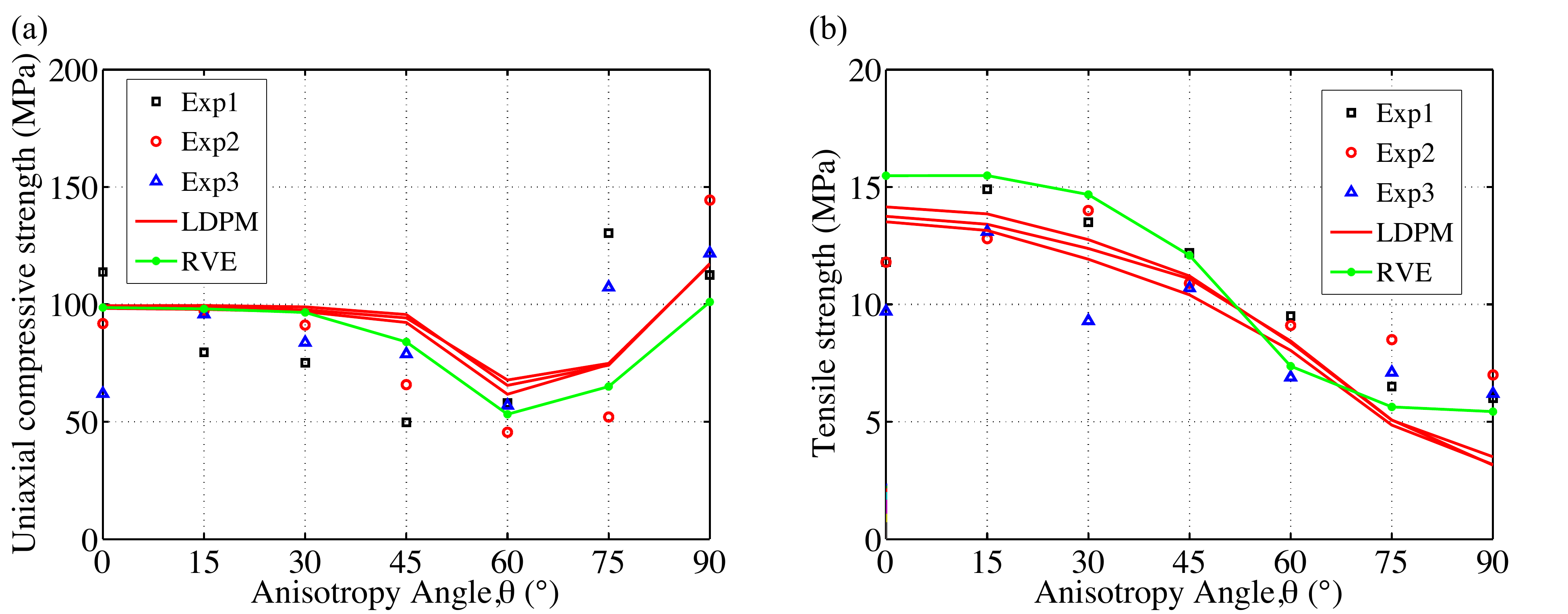}
		\caption{Comparison of (a) uniaxial compressive strength and (b) tensile strength from experiments and simulations as a function of anisotropy angle.}
		\label{fig:UCCBRZCalib}
	\end{figure}

	Figures \ref{fig:UCCBRZCalib}a and b show the variations of UCS and BTS with anisotropy angles respectively. Numerical results (red solid curves) and experimental data (scattered points) are compared, and listed in Table \ref{tab:COMP}. Good agreement can be found between them. 
	
\begin{figure}
		\centering
		\includegraphics[width = 1\textwidth]{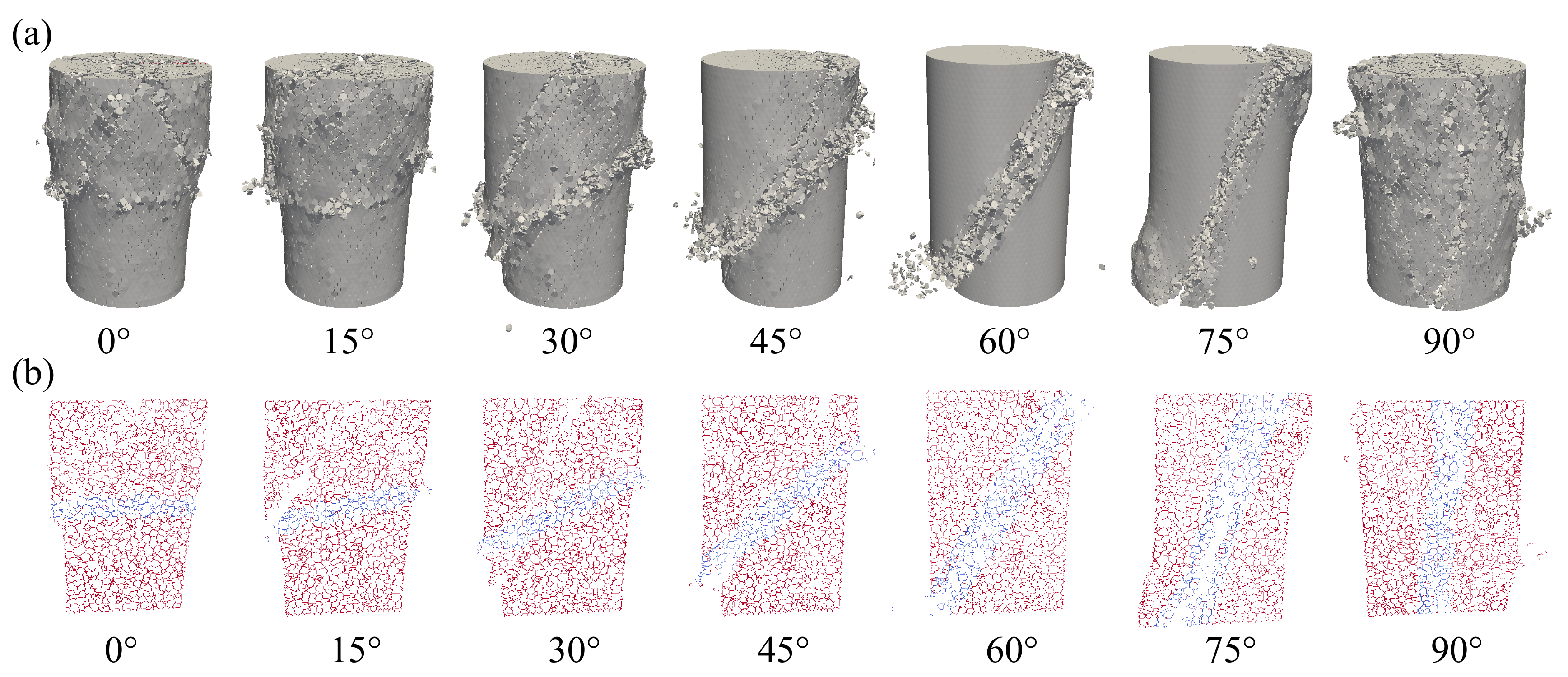}
		\caption{Simulated specimens after failure with anisotropy angles of \ang{0}, \ang{15}, \ang{30}, \ang{45}, \ang{60}, \ang{75} and \ang{90} for uniaxial compression test: (a) snapshot of external faces of simulated specimens; (b) snapshot of slice view of simulated specimens. }
		\label{fig:UCCFailure}
	\end{figure}

	It can be observed from Figure \ref{fig:UCCBRZCalib}a that the maximum UCS occurred at $\theta = \ang{90}$, and the minimum strength occurred at $\ang{45} \leq \theta \leq \ang{75}$. The reduced UCS at certain angles is due to the effect of lamination: when the failure plane coincided with the plane of lamination, the failure occurs at a lower stress level \cite{cho2012deformation}. This effect is mostly governed by $C_{Nt}$, $C_{Ts}$, and $C_{\mu}$, as demonstrated in sections \ref{sec:CNT}, \ref{sec:CTS} and \ref{sec:CMU} respectively. The experimental observation that the measured strength at $\theta = \ang{90}$ is slightly larger than that at $\theta = \ang{0}$ can also be captured by the proposed model. Figure \ref{fig:UCCFailure} shows the simulated specimens with different anisotropy angles after uniaxial compression failure, which is also in good agreement with the experimental results \cite{cho2012deformation}. For specimens with $\ang{45} \leq \theta \leq \ang{75}$, a clear failure plane can be observed almost identical to the plane of lamination, which accounts for the mechanism of the reduced strength at  $\ang{45} \leq \theta \leq \ang{75}$. Note that LDPM does not postulate the existence of a compression failure at the microscopic level, but rather simulates macroscopic compressive failure through tensile and shearing failure at microscale \cite{cusatis2011lattice1}. 
	
\begin{figure}
		\centering
		\includegraphics[width = 1\textwidth]{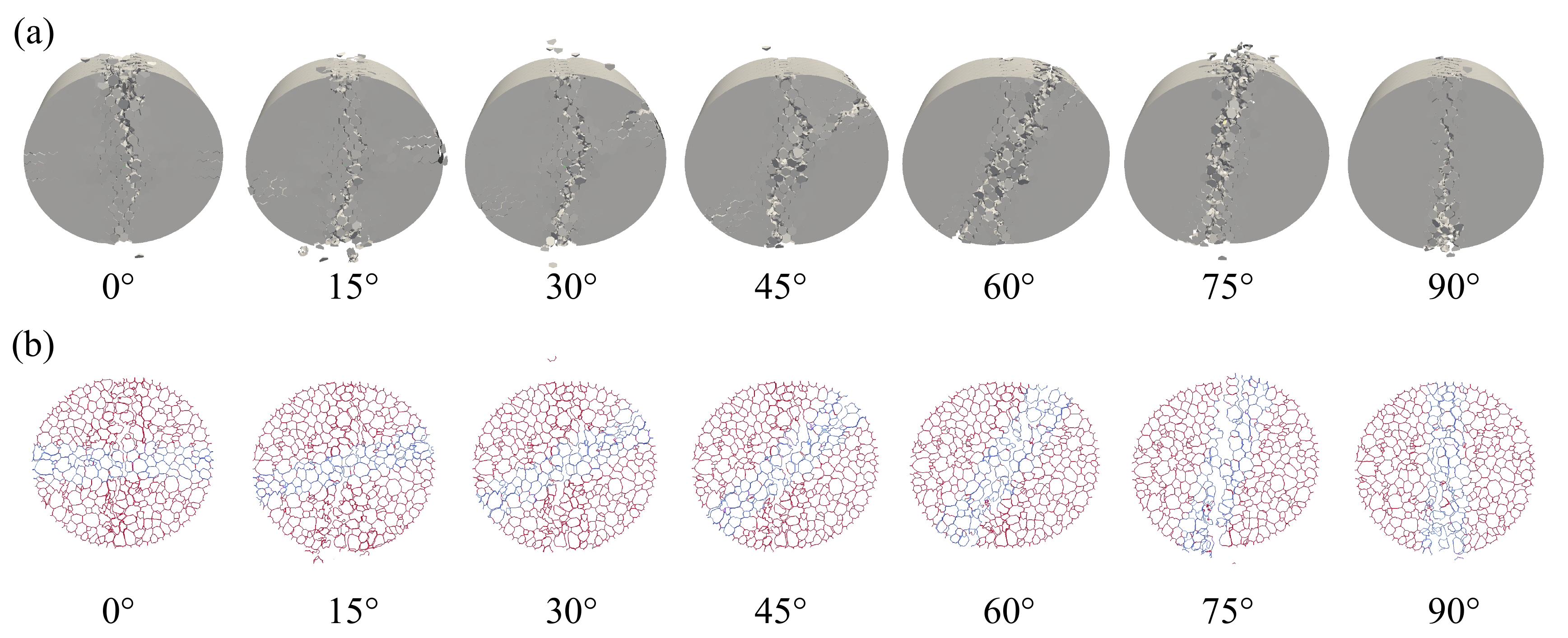}
		\caption{Simulated specimens after failure with anisotropy angles of \ang{0}, \ang{15}, \ang{30}, \ang{45}, \ang{60}, \ang{75} and \ang{90} for Brazilian test: (a) snapshot of external faces of simulated specimens; (b) snapshot of slice view of simulated specimens. }
		\label{fig:BRZFailure}
	\end{figure}

	From Figure \ref{fig:UCCBRZCalib}b, one can find that the maximum macroscopic BTS occurred at $\theta = \ang{0}$, while the minimum value occurred at $\theta = \ang{90}$. Simulated and measured BTS decreases as the anisotropy angle increases. The difference between BTS at $\theta = \ang{0}$ and one at $\theta = \ang{90}$ is governed by LDPM microscale parameters, especially $\sigma_s/\sigma_t$, $C_{Nt}$ and $C_{Ts}$, as demonstrated in section \ref{sec:STratio}, \ref{sec:CNT}, and \ref{sec:CTS}, respectively. From the graph of the simulated specimens at failure, as illustrated in Figure \ref{fig:BRZFailure}, distinctly different failure modes are observed for specimens with different anisotropy angles. For $\ang{0} \leq \theta \leq \ang{15}$, failure develops along the loaded diameter, and the failure mode is purely tensile splitting failure.  For $ \theta \geq \ang{30}$, the failure plane coincides with the plane of lamination. For specimens with anisotropy angles $\ang{30} \leq \theta \leq \ang{75}$, a shear failure plane along lamination can be clearly observed, indicating a shear failure mode mixed with tensile splitting into the matrix; for specimens of $\theta = \ang{90}$, tensile splitting along lamination is dominant. 
	
	To investigate the effect of random microstructure, we also performed calculations of three specimens characterized by different placement of grains, which introduces randomness in the microstructure, for each anisotropy angle. As shown in Figure \ref{fig:UCCBRZCalib}, the calculated UCS and BTS (red solid curves) fall within a narrow scatter band, and the introduced randomness does not alter the trend of UCS and BTS variations. Considering the small deviation resulted from randomness in microstructure, one may conclude that the simulated specimens are large enough to be considered representative. 
	
\begin{figure}
		\centering
		\includegraphics[width = 0.6\textwidth]{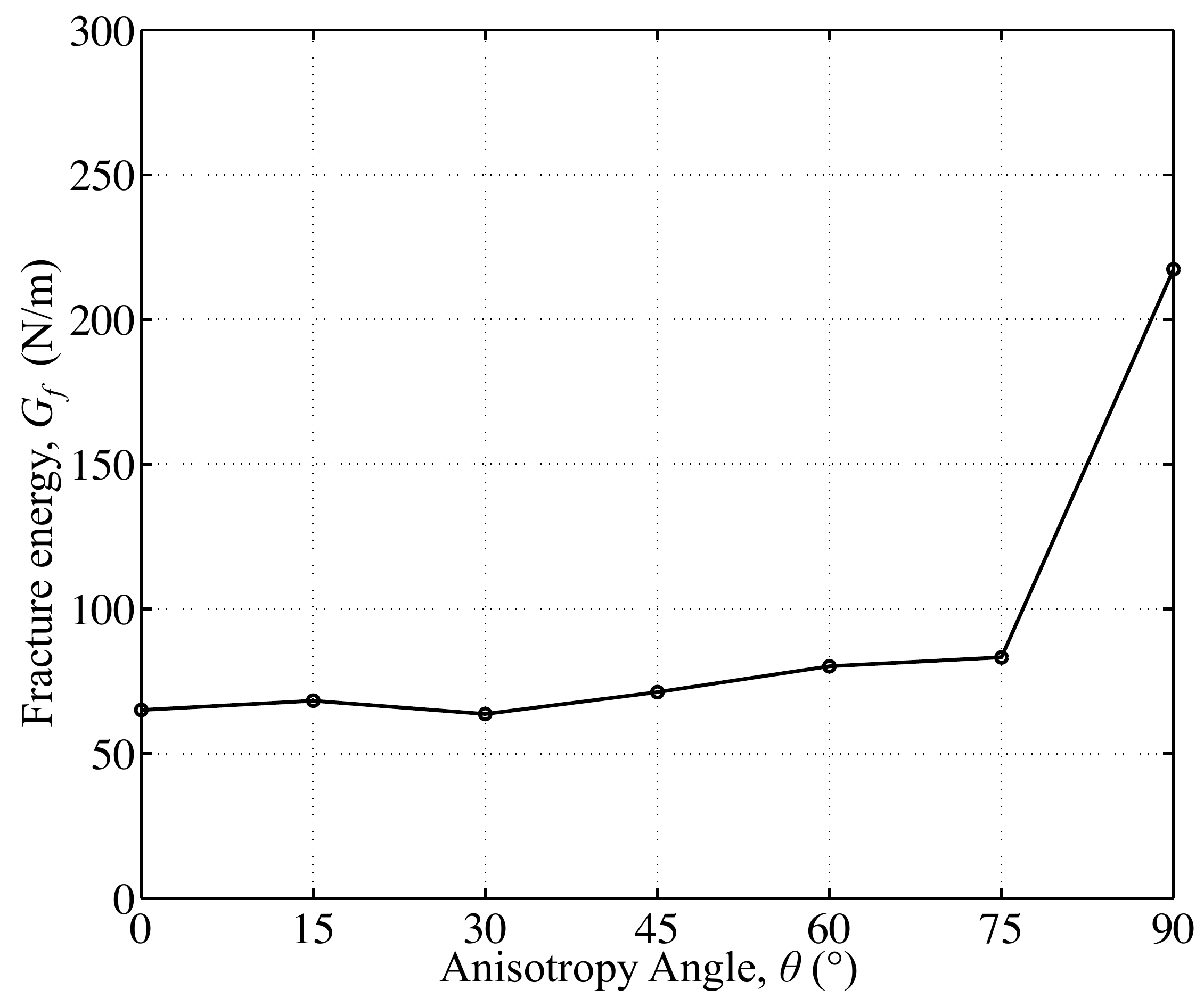}
		\caption{Variation of fracture energy with anisotropy angle. }
		\label{fig:Gf}
	\end{figure}

	Fracture properties of shale are obtained through simulations of uniaxial tensile tests on cylindrical specimens with the same dimension as those of the uniaxial compression tests. Although brittle failure of shale under tensile loading is often observed in experiments conducted at centimeter scale, localization of softening damage into a discrete fracture may be recognized on sufficiently small scale. In this case, the fracture energy, $G_f$, of the material can be deduced from the area $A$ under a complete load-displacement curve with stable postpeak; $G_f = A/A_c$ where $A_c$ is the fractured area of the specimen. Figure \ref{fig:Gf} illustrates the variation of the fracture energy with anisotropy angle calculated in this method and with the model parameters reported in Table \ref{tab:LDPMpara}. The fracture energy of simulated specimen with $\theta = \ang{0}$ is smaller than that with $\theta = \ang{90}$, which agree with the laboratory observation that the fracture toughness measured in the short transverse orientation is smaller than the ones measured in the divider and arrester orientations \cite{chandler2016fracture, schmidt1977fracture, akono2013assessment}. The values of the simulated fracture energy range from 65 N/m to 220 N/m, which are within the same order of magnitude as the work of fracture measured by Chandler et al. \cite{chandler2016fracture} for Mancos shale. The variation of the simulated fracture energy may not be realistic and needs to be calibrated when new experimental data become available.
	
	\begin{table}
		\caption{Values of LDPM parameters used in the simulations.}
		\label{tab:LDPMvalue}
		\centering
		\begin{tabular}{llll}
			\toprule
			Symbol (units)	& Value	& Symbol (units)	& Value\\
			\midrule	
			$E_{N0}$ (GPa)	& 22.5	& $\sigma_{s0}$ (MPa)	& 3.9\\
			$\beta_N$		& 0.24	& $\beta_s$	& 0.13\\
			$C_N$			& 0.4	& $C_{Ts}$	& 0.1\\
			$E_{T0}$ (GPa)	& 4.0	& $l_t$ (mm)& 93\\
			$\beta_T$		& 0.13	& $n_t$		& 0.2\\
			$C_T$			& 0.4	& $\mu_{00}$& 0.2\\
			$\sigma_{t0}$ (MPa)	& 4.8	& $\beta_{\mu}$	& 0.5\\
			$\beta_t$		& 0.24	& $C_{\mu}$	& 0.5\\
			$C_{Nt}$		& 1.0\\
			\bottomrule
		\end{tabular}
	\end{table}			
	
	\begin{table}
		\caption{Values of macroscopic uniaxial compressive strength and tensile strength from simulations and experiments .}
		\label{tab:COMP}
		\centering
		\begin{tabular}{llllllll}
			\toprule
			& \ang{0}	& \ang{15}	& \ang{30}	& \ang{45}	& \ang{60}	& \ang{75}	& \ang{90}\\
			\midrule	
			UCS$_{\text{EXP}}$ (MPa)	& 89.2		& 91.0		& 83.4		& 64.8		& 53.5		& 96.5		& 126.2\\
			UCS$_{\text{NUM}}$ (MPa)	& 99.2		& 98.7		& 97.7		& 94.3		& 67.8		& 74.9		& 117.1\\
			BTS$_{\text{EXP}}$ (MPa)	& 11.1		& 13.6		& 12.3		& 11.3		& 8.5		& 7.4		& 6.4\\
			BTS$_{\text{NUM}}$ (MPa)	& 12.6		& 12.5		& 12.0		& 11.2		& 8.7		& 5.5		& 4.0\\
			\bottomrule
		\end{tabular}
	\end{table}		
	
	\section{Multiscale Homogenization Method}
	The typical grain size of shale is several orders of magnitude smaller than the size of the specimens that are tested in laboratory. For instance, the maximum grain size considered in the above simulations is \SI{50}{\micro\meter}, while the specimen size in Ref. \cite{cho2012deformation} is around \SI{50}{\milli\meter}, three orders of magnitude larger than the maximum grain size. As the proposed model adopts an ``a priori'' discretization and simulates shale at the level of the major heterogeneities (grains in this case), it tends to be computationally expensive, and this hinders its use in the numerical simulations of large systems such as reservoir modeling. In this work, this problem is addressed by a general multiple scale computational theory proposed by Rezakhani and Cusatis \cite{rezakhani2015asymptotic} based on the classical asymptotic expansion homogenization \cite{forest2001asymptotic, chung2001asymptotic, ghosh1996two, fish1997computational}. 
	\subsection{Asymptotic expansion homogenization}	
\begin{figure}
		\centering
		\includegraphics[width = 1\textwidth]{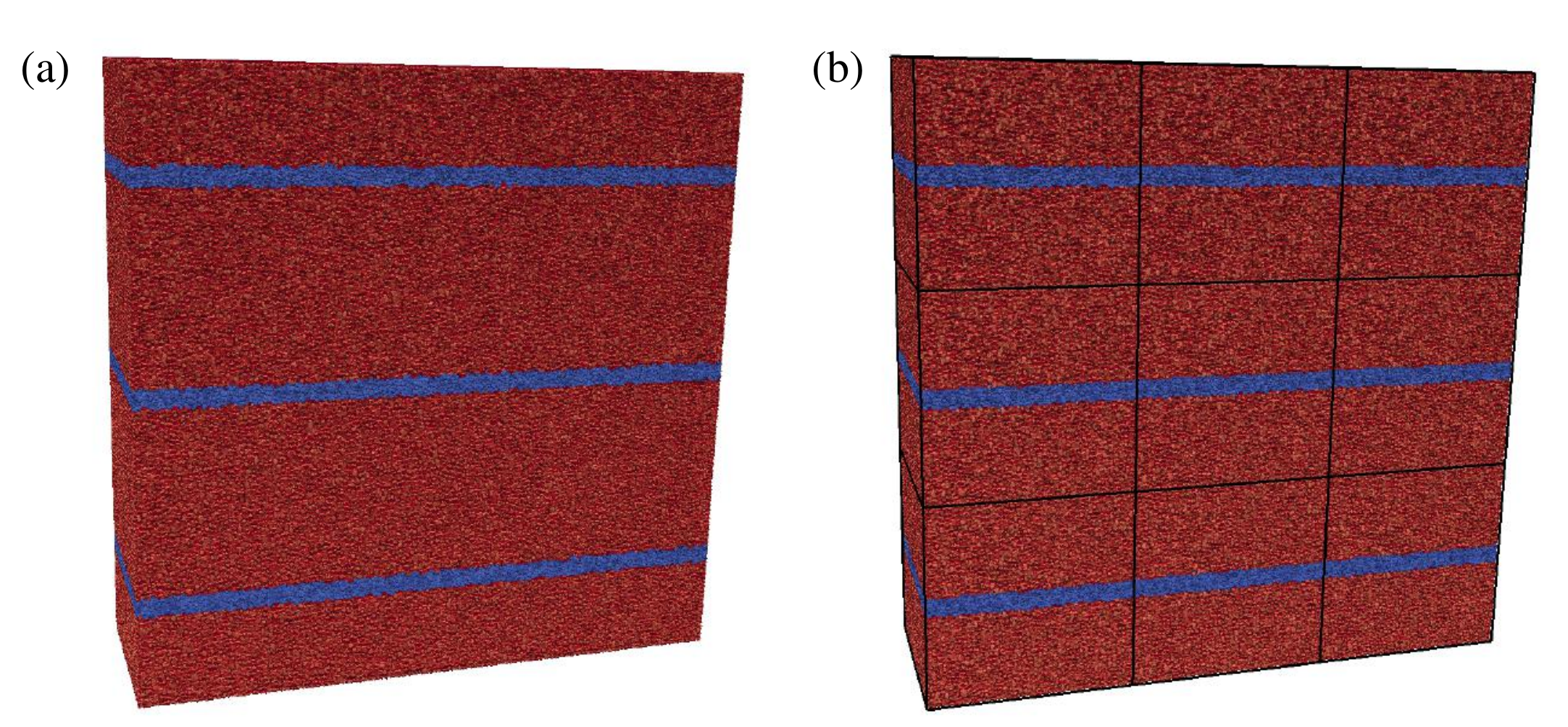}
		\caption{Granular lattice system with lamination in a LDPM prism: (a) generic LDPM system and (b) periodic LDPM system. }
		\label{fig:MHM}
	\end{figure}

	The developed homogenization theory is built on two major assumptions: (1) There exits a certain volume of material, the so called Representative Volume Element (RVE), which properly describes the internal structure of the material under investigation \cite{kouznetsova2004size}; (2) The size of the RVE is much smaller than the characteristic size of the macroscopic problem under consideration, namely the ``separation of scales'' hypothesis holds. Homogenization theory assumes also that the lower-scale material structure is periodic, which means it is composed of a repetition of material RVEs in three dimensions. For the modeling of shale internal structure, polyhedral cells are randomly distributed among the computational domain to represent the cement-coated grains. Fig. \ref{fig:MHM}a shows a typical granular lattice system with lamination generated according to the algorithm discussed in Section \ref{sec:Geom}, and Fig. \ref{fig:MHM}b its periodic approximation. A cube with a single layer is chosen as the RVE in the case of laminated shale, as illustrated in Figure \ref{fig:MHM}b.
	
\begin{figure}[t]
		\centering 
		{\includegraphics[width=\textwidth]{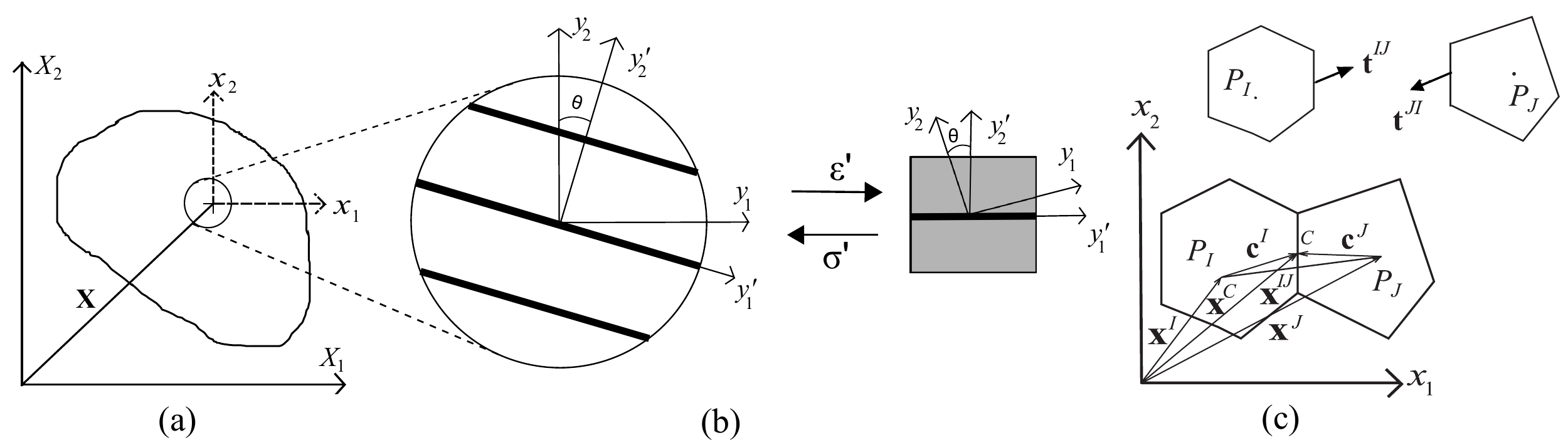}}
		\caption{Geometrical description of the two-scale problem: (a) macro material domain; (b) mesoscale domain with material heterogeneity and Coordinate transformations in the RVE analysis; (c) geometry of two neighboring particles. }
		\label{TwoScaleAnalysis}
	\end{figure}

	A generic macroscopic homogeneous material domain and the corresponding global coordinate system $\bf X$ are depicted in Figure \ref{TwoScaleAnalysis}a. At any point of the macro-scale domain with $\bf X$ position vector, two distinct length scales are introduced to represent (1) the macroscopic problem, in which the material is defined as homogeneous continuum with no detail of material heterogeneity (2) the fine-scale domain, in which heterogeneity is modeled by means of the discrete meso-scale model. $\bf x$ and  $\bf y$, are the local coordinate systems for the macro- and fine-scale problems, respectively. In Figure \ref{TwoScaleAnalysis}b, an enlarged view of the macroscopic material point is illustrated in the local fine-scale coordinate system $\bf{y}$. $\bf{y'}$ is the fine-scale coordinate system which is aligned with the material anisotropy orientation. If the rule of separation of scales holds, macro- and fine-scale coordinate systems are related as    
	\begin{eqnarray}\label{scale-link-1}
		\mathbf{x}=\eta \mathbf{y}  \hspace{0.5 in}   0< \eta <<1
	\end{eqnarray}
	where $\eta$ is a very small positive scalar. Equation \ref{scale-link-1} means that a small distance in the macro-scale coordinate system represents a large distance in the fine-scale space. The homogenization theory starts by considering a generic particle P$_I$ and its interaction with surrounding particles, such as P$_J$, through traction vector as shown in Figure \ref{TwoScaleAnalysis}c. The displacement and rotation of the particle P$_I$,  $\bold{U}^I = \bold{u}(\bold{x}^I, \bold{y}^I)$ and ${\boldsymbol \Theta}^I = \boldsymbol{\theta}(\bold{x}^I, \bold{y}^I)$, can be approximated by means of the following asymptotic expansions
	\begin{equation}
		\bold{u}(\bold x, \bold y) \approx \bold u^0(\bold x, \bold y)+\eta \bold u^1(\bold x, \bold y) 
		\label{disp-expansion}
	\end{equation}
	\begin{equation}
		\boldsymbol{\theta}  (\bold x, \bold y)  \approx \eta^{-1} \boldsymbol{\omega} ^{0}(\bold x, \bold y) +  \boldsymbol{\varphi}^0(\bold x, \bold y)+  \boldsymbol{\omega}^{1}(\bold x,\bold y)+\eta \boldsymbol{\varphi}^{1}(\bold x, \bold y) 
		\label{rot-expansion}
	\end{equation}
	where terms up to order $\mathcal{O}(\eta)$ are considered. $\bold  u^0(\bold x, \bold y)$, and $\bold  u^1(\bold x, \bold y)$ are respectively the macro- and the fine-scale displacement fields. Asymptotic expansion of rotation field is written considering the fact that rotation vector corresponds to the curl of displacement vector. Therefore, $\boldsymbol{\omega}^{0}$, $\boldsymbol{\omega}^{1}$ are the rotations in the fine-scale space, while $\boldsymbol{\varphi}^{0}$, $\boldsymbol{\varphi}^{1}$ are the corresponding coarse-scale ones. One should consider that, contrarily to the expansion of displacements, the asymptotic expansion for rotations features a term of order $\mathcal{O}(\eta^{-1})$  and two distinct terms of order $\mathcal{O}(1)$. Substituting Equations \ref{disp-expansion} and \ref{rot-expansion} into the definition of facet strains, Equation \ref{ldpm-strain}, and using the macroscopic Taylor expansion of displacement and rotation of node $P_J$ around node $P_I$, the following form for the multiple scale definition of facet strains is obtained \cite{rezakhani2015asymptotic}
	\begin{equation}\label{eps-expansion}
		\epsilon_{\alpha}=\eta^{-1} \epsilon_{\alpha}^{-1} + \epsilon_{\alpha}^0 + \eta \epsilon_{\alpha}^1
	\end{equation}
	where $\epsilon_{\alpha}=$ facet strains; $\alpha = N, M, L$ with $N$ representing normal components and $M,L$ representating tangential components. Considering Eq. \ref{eps-expansion} and the facet constitutive equations, it is shown in Ref. \cite{rezakhani2015asymptotic} that the multiple scale definition of facet tractions is $t_{\alpha} = \eta^{-1} t^{-1}_{\alpha} +  t^{0}_{\alpha} + \eta t^{1}_{\alpha}$, which is assumed to be valid in both elastic and nonlinear regime. Substituting the asymptotic expansion of facet tractions in the particle equilibrium equations (see Eq. \ref{motion-1} and \ref{motion-2}), one can derive separate scale governing equations for both fine- and coarse-scale problems. After some analytical derivation, the final governing equations for the RVE and macroscopic problems are obtained as presented below (for details on the calculations see Ref. \cite{rezakhani2015asymptotic}). 
	
	\subsubsection{Fine-scale equation: the RVE problem}
	The RVE problem is governed by the following equations which are force and moment equilibrium equations of each particle inside the RVE
	\begin{equation}\label{RVE-1}
		\sum_{\mathcal{F}_I}{{A}\, t^{0}_{\alpha} {\boldsymbol{e}}_\alpha^{IJ}} = 0  \hspace{0.4 in} \sum_{\mathcal{F}_I} {A}\, (\boldsymbol{c}^I \times t^{0}_{\alpha} \boldsymbol{e}_\alpha^{IJ}) = 0
	\end{equation}
	Facet traction vectors of $\mathcal{O}(1)$, $t_\alpha^0$, which appeared in above equation are functions of $\epsilon^0_\alpha$ that is derived as
	\begin{equation}\label{eps-expansion-zero'}
		\epsilon_{\alpha}^0 = {r}^{-1} \left( u_i^{1J} - u^{1I}_i + \varepsilon_{ijk}  \omega_j^{1J} c_{k}^{J} - \varepsilon_{ijk} \omega_j^{1I} c_{k}^{I}\right) {e}^{IJ}_{\alpha i} + P^\alpha_{ij} \left(\gamma_{ij}  
		+ \varepsilon_{jmn}  \kappa_{im} y_{n}^{c} \right) 
	\end{equation}
	where $\gamma_{ij} = v^{0}_{j,i} - \varepsilon_{ijk} \omega_k^{0}$ and $\kappa_{ij}=\omega_{j,i}^{0}$ are the macroscopic Cosserat strain and curvature tensors, respectively. $r=|\mathbf{x}^{IJ}|$ is the length of the vector $\mathbf{x}^{IJ}$ which connects the mass centers of particles the P$_I$ and P$_J$, as shown in Figure \ref{TwoScaleAnalysis}c. The vector $\boldsymbol{y}^c$ is the vector that connects the centroid of the facet shared between particles $I$ and $J$ to the mass center of the RVE. $P^\alpha_{ij} = n^{IJ}_i e^{IJ}_{\alpha j}$ is a projection operator. The first term of Eq. \ref{eps-expansion-zero'} is the definition of the facet strains (one normal and two tangential) written in terms of fine-scale displacements and rotations $u^1$ and $\omega^1$. The second term of Equation \ref{eps-expansion-zero'}, $P^\alpha_{ij} \left(\gamma_{ij} + \varepsilon_{jmn}  \kappa_{im} y_{n}^{c} \right)$, is the projection of macroscopic Cosserat strain and curvature tensors on each facet. Therefore, the $\mathcal{O}(1)$ facet strains is the sum of their fine-scale counterparts and the projection of the macroscopic strain and curvature tensors onto the facet level. In other words, strain and curvature tensors at each macroscopic computational point are applied on the associated RVE as imposed negative eigen-strains on all its facets. The RVE is analyzed under periodic boundary conditions, and it leads to the calculation of the fine-scale quantities $\mathbf{u}^1$ and $\boldsymbol{\omega}^1$. It should be noted that the fine-scale length type variables in Equation \ref{eps-expansion-zero'}, $r$, $c_k^I$, $c_k^J$, and $y_n^c$,  must be measured in the fine-scale coordinate system.
	
	\subsubsection{Coarse-scale equation: the macroscopic problem}
	Mathematical manipulation of the $\mathcal{O}(1)$ terms in the multiple scale expansion of particle equilibrium equations lead to the  macroscopic translational and rotational equilibrium equations. By averaging the equilibrium equations over all RVE particles, the macro-scale translational equilibrium equation and the corresponding homogenized stress tensor are expressed as 
	\begin{equation} \label{macro-eq-cont}
		\sigma^0_{ji,j} + b_i = 0
	\end{equation}
	\begin{equation} \label{macro-stress-formula}
		\sigma^0_{ij} = \frac{1}{2V_0} \sum_I \sum_{\mathcal{F}_I}{A} r t^0_\alpha P_{i j}^{\alpha}
	\end{equation}
	where $V_0$ is the volume of the RVE; $\rho_u=\sum_I {M}^I_u/V_0$ is the mass density of the macroscopic continuum. Equation \ref{macro-eq-cont} is the classical partial differential equation governing the equilibrium of continua whereas Equation \ref{macro-stress-formula} provides the macroscopic stress tensor though homogenizing the solution of the RVE problem. In addition, the final macro-scale rotational equilibrium equation and the corresponding macroscopic moment stress tensor are derived as
	\begin{equation} \label{macro-rotational-final} 
		\begin{gathered}
			{\epsilon}_{ijk} {\sigma}_{ij}^0 + \frac{\partial \mu^0_{ji}}{\partial x_j} = 0
		\end{gathered}
	\end{equation}
	\begin{equation} \label{macro-momentstress-formula} 
		\begin{gathered}
			\mu^0_{ij} = \frac{1}{2V_0}\sum_I \sum_{\mathcal{F}_I} {A}r t_{\alpha}^0 Q_{ij}^{\alpha}
		\end{gathered}
	\end{equation}
	where the projection matrix $Q_{ij}^{\alpha}$ is defined as $Q_{ij}^{\alpha} = n_i^{IJ} \varepsilon_{jkl} x^C_k e_{\alpha l}^{IJ}$. $\mu^0_{ij}$ is the macroscopic moment stress tensor derived based on the results of the RVE analysis, and Equation \ref{macro-rotational-final} corresponds to the classical rotational equilibrium equation in Cosserat continuum theory. One can find the derivation details of above equations in Ref. \cite{rezakhani2015asymptotic}. The presented homogenization theory has been used recently to obtain macroscopic properties of concrete RVE simulated by LDPM and to homogenize concrete elastic and nonlinear behavior under different types of loading conditions, see Refs.  \cite{rezakhani2015asymptotic,cusatis2014multiscale}.
	
	\subsubsection{RVE with material anisotropy}
	As presented in the previous sections, the material anisotropy of shale is modeled through an approximated geometric description of shale internal structure and a representation of material properties variation. To generate a RVE with anisotropy angle $\theta$, a method of coordinate transformation is presented in this section. 
	
	Let us consider the zoomed view of the macroscopic material point with the local fine-scale coordinate system  $\bf y$ as shown in Figure \ref{TwoScaleAnalysis}b and let us suppose that the normal vector of the weak layers, which is parallel to the axis of symmetry, has a angle $\theta$ with respect to the $y_2$ axis. The angle $\theta$ is the anisotropy angle defined in the previous sections. One can define a local coordinate system $\bf y'$ anchored to a RVE as shown in Figure \ref{TwoScaleAnalysis}d with $y'_2$ parallel and $y'_1$, $y'_3$ perpendicular to the normal vector. The coordinate system $\bf y'$ can be considered as being rotated counter-clockwise by the angle $\theta$ from the coordinate system $\bf y$. A transformation of a 2nd rank tensor $F_{ij}$ from the coordinate system $\bf y$ to tensor $F'_{ij}$ in the coordinate system $\bf y'$ is conducted as follows
	\begin{equation}
		\label{eqn:CoordTrans}
		F'_{ij} = R_{ip} R_{jq} F_{pq}
	\end{equation} 
	where the components of transformation matrix $\boldsymbol{R}$ can be written as
	\begin{equation}
		\boldsymbol{R} = 
		\begin{bmatrix}
			\cos \theta 	& \sin \theta & 0 \\
			-\sin \theta 	& \cos \theta & 0 \\
			0 & 0 & 1
		\end{bmatrix}
	\end{equation}
	In this case, the tensor $F_{ij}$ can be macroscopic Cosserat strain tensor $\gamma_{ij}$, curvature tensor $\kappa_{ij}$, stress tensor $\sigma_{ij}^0$, and moment stress tensor $\mu_{ij}^0$. 
	
	\subsection{Numerical results}
	The homogenization theory formulated above was implemented in the MARS computational software with the objective of upscaling the shale fine-scale model discussed above. The procedure to construct a generic RVE is described in Ref. \cite{rezakhani2015asymptotic}. The RVE analysis is conducted by imposing periodic boundary conditions. This is obtained by setting the displacements and rotations of the RVE vertexes to be zeros and by imposing that the periodic edge nodes and face nodes have the same rotations and displacements. 
	
	In general, the overall multiscale numerical procedure adopted in this paper is similar to the one proposed in Ref. \cite{rezakhani2015asymptotic}. However, in order to accommodate material anisotropy, the procedure is modified and summarized as follows.	
	
	\begin{enumerate}
		\item[(i)] The finite element method is employed to solve the macroscale homogeneous problem in which external loads and essential BCs are applied incrementally. During each numerical step, strain increments $\Delta \gamma_{ij} = \Delta v^0_{j,i} - \varepsilon_{ijk} \Delta \varphi^0_k$ and curvature increments $\Delta \kappa_{ij} = \Delta \omega^0_{j,i}$ tensors are calculated at each integration point based on the nodal displacement and rotation increments of the corresponding finite element. 
		\item[(ii)] The macroscopic Cosserat strain and curvature increments $\Delta \gamma_{ij}$ and $\Delta \kappa_{ij}$ are transformed to $\Delta \gamma'_{ij}$ and $\Delta \kappa'_{ij}$ in the coordinate system $\bf y'$ through Equation \ref{eqn:CoordTrans}. 
		\item[(iii)] $\Delta \gamma'_{ij}$ and $\Delta \kappa'_{ij}$ are projected into the RVE facets through the proper projection operations. These projected strains and curvatures are imposed to the RVE allowing the calculation of the fine-scale solution governed by the fine-scale constitutive equations. 
		\item[(iv)] The fine-scale facet tractions are used to compute the macroscopic stress $\sigma'^0_{ij}$, and couple stresses, $\mu'^0_{ij}$, for each Gauss point in the FE mesh.  
		\item[(iv)] The tensors $\sigma_{ij}^0$ and $\mu_{ij}^0$ in the original coordinate system $\bf y$ are obtained through the transformation of the tensors  $\sigma'^0_{ij}$ and $\mu'^0_{ij}$. 	
	\end{enumerate}
	It was shown in Ref. \cite{rezakhani2015asymptotic} that the coarse-scale couple-curvature constitutive equations scale with the square of the RVE size in both linear elastic and nonlinear ranges. Since the size of RVEs is usually smaller than the size of finite elements, it is necessary to accommodate the size dependency of moment stress in the procedure of performing multiscale analysis.  
	
	In a general FEM framework, the macroscopic Cosserat strain and curvature are interpolated by their values at Gaussian integration points. To tackle the size dependency of moment stress, the curvature increments after coordinate transformations at each integration point, i.e. $\Delta \kappa'^{\text{(IP)}}_{ij}$, are scaled by a factor $D^{\text{(IP)}}/D^{\text{(UC)}}$, i.e.  $\Delta \kappa'^{\text{(Eff)}}_{ij} =  \Delta \kappa'^{\text{(IP)}}_{ij} D^{\text{(IP)}}/D^{\text{(UC)}}$, where $D^{\text{(UC)}}$ is the size of RVE, $D^{\text{(IP)}} = (V^{\text{(IP)}})^{1/3}$ where $V^{(IP)}$ is the volume associated to an integration point. The effective curvature increments $\Delta \kappa'^{\text{(Eff)}}_{ij}$ are then projected into the RVE facets in step (iii). In step (iv), the effective macroscopic couple stresses $\mu'^{0\text{(Eff)}}_{ij}$ obtained by RVE calculations are scaled by the same factor, i.e. $\mu'^{0\text{(IP)}}_{ij} = \mu'^{0\text{(Eff)}}_{ij} D^{\text{(IP)}}/D^{\text{(UC)}}$ to get the macroscopic couple stresses $\mu'^{0\text{(IP)}}_{ij}$ for each integration point.

	\subsubsection{Elastic RVE analysis}
	This section presents the analysis of the elastic macroscopic behavior of one LDPM RVE. Rezakhani and Cusatis \cite{rezakhani2015asymptotic} concluded that the macroscale elastic parameters relating the stress tensor to the strain tensor become independent on RVE size, and on the random position of the polyhedral particles inside the RVE, as long as the RVE size is larger than about five times the maximum spherical particle size in the framework of LDPM. To verify this statement, we performed numerical calculations with varying RVE sizes. Although the LDPM RVE size is determined by the spacing of weak layers once lamination is introduced, the elastic analysis is first performed on RVEs without lamination to isolate the effect of grain size compared to the one of RVE size. Five RVE sizes, $D = 0.10,0.15,0.25,0.50, \text{and } 1.00$ \SI{}{mm}, are considered, and five RVEs, characterized by different placement of grains, are studied for each case. In LDPM, different placement of grains inside the RVE yields to distinct RVE polyhedral particle configurations. The calculations are performed by assuming the LDPM elastic parameters reported in the last row of Table \ref{tab:ElastLDPM}.
	
\begin{figure}
		\centering
		\includegraphics[width = 1\textwidth]{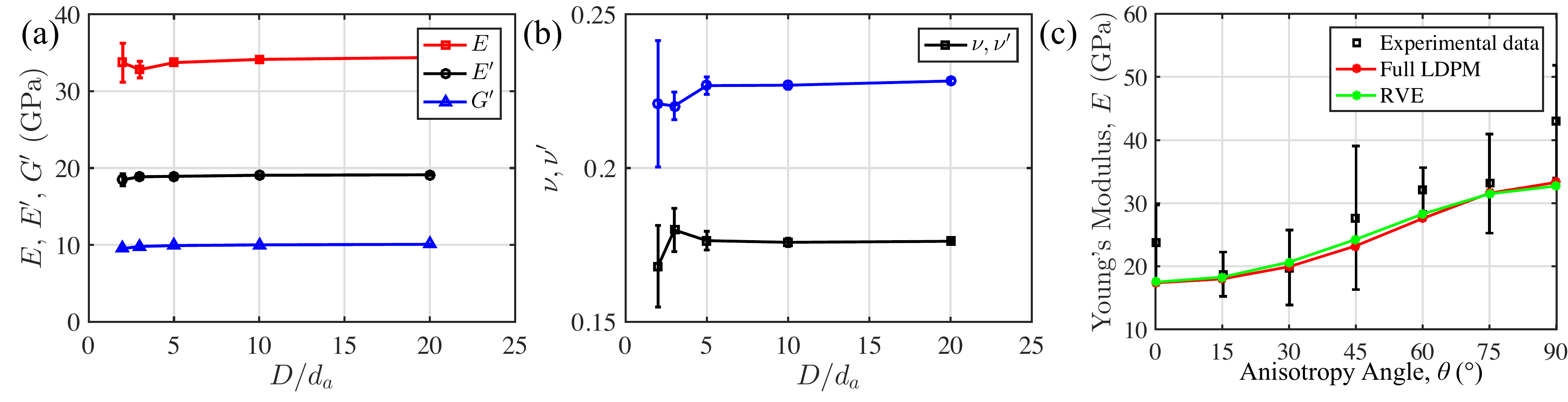}
		\caption{Variation of elastic homogenized material properties with respect to the ratio of RVE size to maximum particle size: (a) Young moduli $E$, $E'$ and shear modulus $G'$; (b) Poisson's ratios $\nu$ and $\nu'$. (c) Variation of in-plane Young's modulus with anisotropy angles. }
		\label{fig:UCNL}	
	\end{figure}

	Figure \ref{fig:UCNL} shows the homogenized values of the five elastic constants $E, E', G', \nu$ and $\nu'$ as functions of the RVE size normalized by the maximum particle size, $d =  D/d_a$. The error bars represent the scatter in the results obtained by simulating five different RVEs of the same size but with different realization of grain positions inside the RVE. The calculated values of the parameters tend to converge to a constant value as the size of the RVE increases. The homogenized values of these elastic properties become independent of the grain distribution inside the RVE for value of $d = D/d_a$ greater than 5, as already verified in \cite{rezakhani2015asymptotic} for isotropic materials. 
	
	The RVE with lamination has a fixed size $D = \SI{1.00}{mm}$. The elastic RVE analysis were performed on seven laminated RVEs with anisotropy angles ranging from \ang{0} to \ang{90} respectively. The homogenized values of the Young's modulus in the plane of transversely isotropy $E$ were calculated. As shown in Figure \ref{fig:UCNL}c, the variation of homogenized $E$ with anisotropy angles $\theta$ agrees well with the results of the full LDPM analysis in section \ref{sec:elasticanalysis} and experimental data \cite{cho2012deformation}. 

	\subsubsection{Tension and compression tests on a single tetrahedral element}
\begin{figure}
		\centering
		\includegraphics[width = 1.0\textwidth]{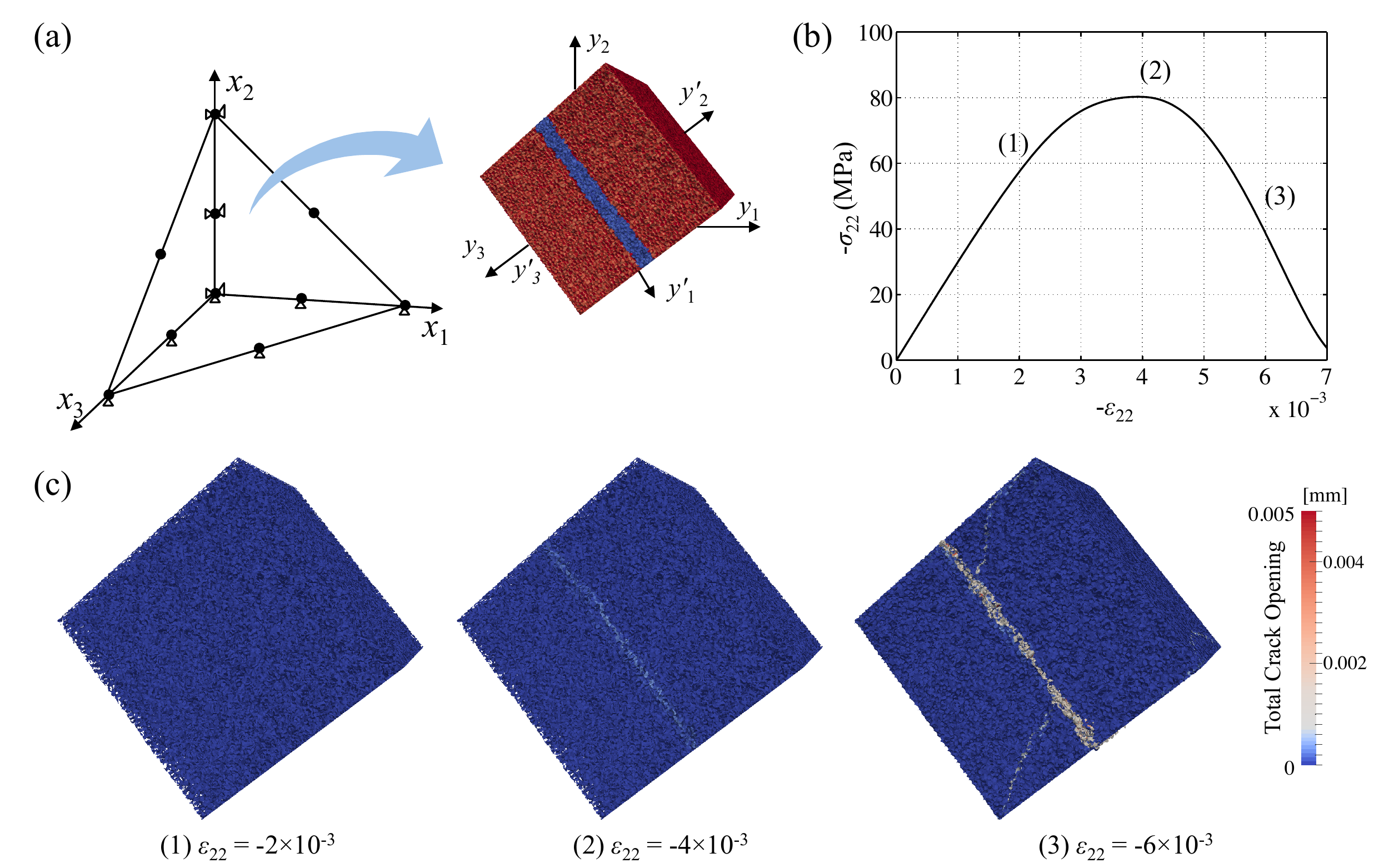}
		\caption{(a) Analysis of a tetrahedral element through a two-scale homogenization algorithm. (b)-(c) Total crack opening evolution of a RVE with the anisotropy angle of \ang{75} in single element compression test.}
		\label{fig:TetSingle}	
	\end{figure}

	The behavior of a single tetrahedral element (Fig. \ref{fig:TetSingle}a) under uniaxial tension and compression loading is studied through a two-scale homogenization algorithm,  to investigate the significance of the homogenized macroscopic moment stress tensor. The obtained results are also compared to these from the simulated unconfined compression and Brazilian tests described in the previous sections. A 10-nodes tetrahedral finite element with three translational and three rotational degrees of freedom \cite{zhou2014tetrahedral} is considered, and a RVE is assigned to every macroscopic integration point. Figure \ref{fig:TetSingle}a depicts the boundary condition applied on the tetrahedral element. The top vertex is pulled along the $x_2$-direction up to a displacement equal to \SI{0.01}{mm} in the tension test and is pushed down to \SI{0.04}{mm} in the compression test. All rotational degrees of freedom are released to allow the development of the macroscopic Cosserat curvature tensors during the loading process. The edge length of the element is chosen to be \SI{3}{mm} as against the RVE size of \SI{1}{mm}. The same as previous sections, seven different anisotropy angles are considered herein. The geometric parameters and the LDPM material parameters are the same as those used in the Brazilian and uniaxial compression tests, as reported in Table \ref{tab:LDPMvalue}.
	
	\begin{figure}
		\centering
		\includegraphics[width = 1.0\textwidth]{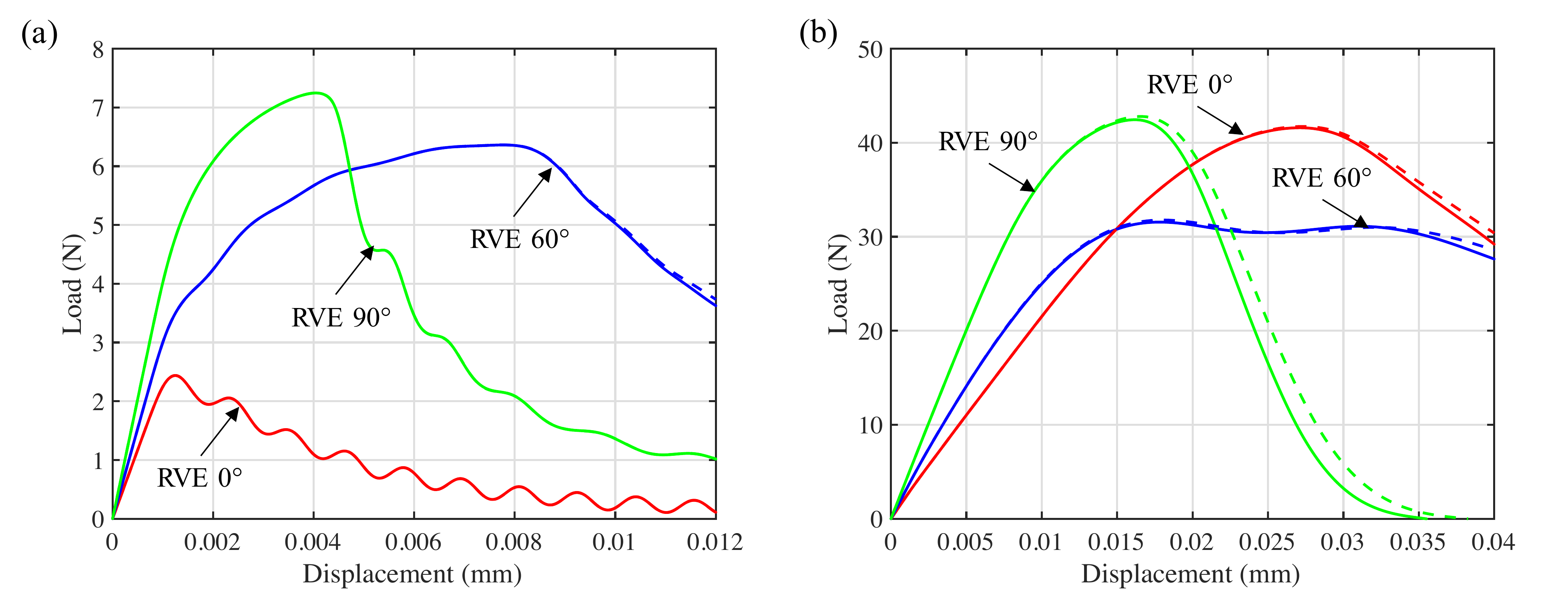}
		\caption{Comparison of simulated load-displacement curves of single tetrahedral element through the two-scale homogenization algorithm with and without considering the macroscopic moment tensor represented by solid and dash lines respectively: (a) uniaxial tension test and (b) uniaxial compression test.}
		\label{fig:TetMomStress}	
	\end{figure}			
	
	Figure \ref{fig:TetMomStress} compares the responses of the tetrahedral element tests through the two-scale homogenization algorithm with and without considering the macroscopic moment tensor calculated by Eq. \ref{macro-momentstress-formula}. The load-displacement curves obtained by simulated tension and compression tests with three RVEs of anisotropy angles \ang{0}, \ang{60}, and \ang{90} respectively are presented. As one can observe from Figure \ref{fig:TetMomStress}a, the responses of the tetrahedral element under uniaxial tension with and without considering the moment tensor are almost identical in each case with RVEs of different anisotropy angles. A slight difference of element responses in the post-peak region during the compression test can be observed in Figure \ref{fig:TetMomStress}b. This is a result of layer slippage and the development of the shear failure plane, which may lead to the violation of stress symmetry \cite{adhikary1997cosserat}. Generally, the effect of the macroscopic moment tensor is negligible in both tension and compression tests, especially in this paper in which only the pre-peak region was calibrated for the lack of post-peak experimental data.

	The homogenized values of the tensile and compressive strengths of the RVEs are obtained by the macroscopic $\sigma_{22} - \varepsilon_{22}$ curves of one integration point in the tetrahedral element. The variations of uniaxial compressive strength with anisotropy angles calculated from the RVE response of two-sale homogenization algorithm are shown in Figure \ref{fig:UCCBRZCalib}a, and compared with the results of the uniaxial compression test through full LDPM simulations in section \ref{sec:Calib} and the experimental data reported in \cite{cho2012deformation}. Good agreement between them can be observed. Similarly, Figure \ref{fig:UCCBRZCalib}b compares the tensile strength obtained by RVE responses, full LDPM analysis, and experimental data. Note that the simulated RVE responses in this section give the direct tensile strength, while the full LDPM simulations and experimental data give the indirect tensile strength (Brazilian tensile strength). Although a small discrepancy between the calculated direct and indirect tensile strength can be observed, agreement between RVE and full LDPM results is overall satisfactory. 
	
	Evolution of total crack opening for a RVE with anisotropy angle of \ang{75} during the element compression test is also shown in Figure \ref{fig:TetSingle}b-c at three different macroscopic strain, $\varepsilon_{22}$, levels. Strain level (1) is in pre-peak regime in which damage is distributed through the RVE; at strain level (2) which corresponds to the peak of the stress-strain curve, damage begins to distributed locally along the plane of lamination; strain level (3) is relevant to the post-peak regime in which localization of crack opening occurs following the development of shear failure plane. The observed failure mode in the RVE is similar to the one in the full LDPM analysis shown in Figure \ref{fig:UCCFailure}.
	
	\section{Conclusion}
	A discrete micromechanincal approach based on the Lattice Discrete Particle Model (LDPM) and a multiscale framework based on the asymptotic expansion homogenization theory were formulated and calibrated to simulate the anisotropic mechanical behavior of shale. Based on the results presented in this work, the following conclusion can be drawn. 
	
	\begin{enumerate}
		\item[$\bullet$] 
		The proposed model can simulate the mechanical behavior of anisotropic shale featuring the orientational dependence of the elastic stiffness and strength. Benefiting from the framework of discrete models and the ``a priori'' discretization approach adopted by LDPM, the model addresses material anisotropy through an approximated geometric description of shale internal structure, which includes representation of material property variation with orientation within the constitutive equations and explicit modeling of lamination. It reproduces successfully the variations of Young's modulus, uniaxial tensile and compressive strengths with different bedding plane orientations in the simulated compression and Brazilian tests. 
		
		\item[$\bullet$]  
		The model also succeeds in capturing different failure mechanisms in the simulated experiments caused by varying orientations of shale bedding planes with respect to loading direction. It was shown that distinctly different failure modes can be observed in the simulated uniaxial compression and Brazilian tests, which contributes to the directional dependence of the simulated tensile and compressive strengths respectively. 
		
		\item[$\bullet$]  
		The model is able to simulate the fracture behavior of anisotropic shale under tensile loading. It needs to be furthered calibrated and verified once a relevant experimental study becomes available. 
		
		\item[$\bullet$]  
		The multiscale structure of shale is addressed numerically by the proposed multiscale computational framework. The developed micromechanical based model is upscaled to approximate effective material characteristics at the macroscopic level through a mathematical homogenization approach. The equivalent homogenized continuum is of Cosserat-type, and the material anisotropy of shale is tackled at the RVE level. 
		
		\item[$\bullet$] 
		Similar to previous research on isotropic materials, the macroscale elastic parameters relating the stress to strain tensors become independent of RVE size and the random position of the grains inside the RVE for the ratio of RVE sizes to the maximum particle size greater than 5.  
		
		\item[$\bullet$]  
		The homogenization algorithm was further verified by the simulations of uniaxial tension and compression tests on a single tetrahedral element. Although the nonsymmetric part of stress tensors featured by the Cosserat continuum formulation may arise as a result of layer slippage in the post-peak region, the numerical results indicate the insignificance of the effect of the macroscopic moment tensor. 
		
	\end{enumerate}

	
	\section*{ACKNOWLEDGEMENTS}
	The authors would like to acknowledge the Institute for Sustainability and Energy at Northwestern
	(ISEN) funding scheme. This research was also supported by the Center for Sustainable Engineering of Geological and Infrastructure Materials (SEGIM) and the Quest high performance computing facility at Northwestern University.

	\bibliography{bib1}
	\bibliographystyle{wileyj}

\end{document}